\documentclass[journal]{IEEEtran}

\usepackage{graphicx}
\usepackage{amsmath}
\DeclareUnicodeCharacter{2212}{-}
\ifCLASSINFOpdf
\else
\fi

\usepackage{amsmath}
\usepackage{caption}
\usepackage{bm}
\usepackage{xcolor}
\usepackage[normalem]{ulem}  
\usepackage{makecell} 

\def\beq{\begin{equation}}
\def\eeq{\end{equation}}
\newcommand{\bs}[1]{\ensuremath{\boldsymbol{#1}}}
\def\I{{\mathbf{I}}}

\def\bu{\bs{u}}
\def\bx{\bs{x}}

\begin{document}
%
\title{2D Ultrasound Elasticity Imaging of Abdominal Aortic Aneurysms Using Deep Neural Networks}
%
%
%
\author{
    Utsav Ratna Tuladhar$^{1}$, Richard Simon$^{1}$, Doran Mix$^{2}$, Michael Richards$^{1,2}$ \\
    \IEEEauthorblockA{$^{1}$Rochester Institute of Technology, Rochester, NY, USA} \\
    \IEEEauthorblockA{$^{2}$University of Rochester Medical Center, Rochester, NY, USA} \\
    Email: \{ut3320, rasbme, msrbme\}@rit.edu, doran.mix@urmc.rochester.edu
}

\markboth{Journal of \LaTeX\ Class Files,~Vol.~14, No.~8, August~2015}%
{Shell \MakeLowercase{\textit{et al.}}: Bare Demo of IEEEtran.cls for IEEE Journals}

%



\maketitle

\begin{abstract}

Abdominal aortic aneurysms (AAA) pose a significant clinical risk due to their potential for rupture, which is often asymptomatic but can be fatal. Although maximum diameter is commonly used for risk assessment, diameter alone is insufficient as it does not capture the properties of the underlying material of the vessel wall, which play a critical role in determining the risk of rupture. To overcome this limitation, we propose a deep learning-based framework for elasticity imaging of AAAs with 2D ultrasound. Leveraging finite element simulations, we generate a diverse dataset of displacement fields with their corresponding modulus distributions. We train a model with U-Net architecture and normalized mean squared error (NMSE) to infer the spatial modulus distribution from the axial and lateral components of the displacement fields. This model is evaluated across three experimental domains: digital phantom data from 3D COMSOL simulations, physical phantom experiments using biomechanically distinct vessel models, and clinical ultrasound exams from AAA patients. 
Our simulated results demonstrate that the proposed deep learning model is able to reconstruct modulus distributions, achieving an NMSE score of 0.73\%. Similarly, in phantom data, the predicted modular ratio closely matches the expected values, affirming the model's ability to generalize to phantom data.
We compare our approach with an iterative method which shows comparable performance but higher computation time. In contrast, the deep learning method can provide quick and effective estimates of tissue stiffness from ultrasound images, which could help assess the risk of AAA rupture without invasive procedures.

\end{abstract}


%
\IEEEpeerreviewmaketitle

\section{Introduction}
Abdominal aortic aneurysm (AAA) is a life-threatening cardiovascular disease characterized by localized enlargement and weakening of the abdominal aorta. Left untreated, the progression of an aneurysm can lead to its rupture, a catastrophic event associated with a high mortality rate \cite{aggarwal2011abdominal}. In addition, AAAs are often asymptomatic, highlighting the importance of early detection and continuous monitoring \cite{pande2008abdominal}. Most AAAs are identified incidentally through screening programs or diagnostic imaging for unrelated conditions. Individuals with certain risk factors \cite{kent2010analysis}, such as old age, male sex, smoking history, or a family history of aneurysms, are at increased risk and therefore are prioritized for screening \cite{fleming2005screening}. For such screening, ultrasound imaging is a particularly reliable and preferred method \cite{{benson2018ultrasound},{lindholt1999validity}} due to its accurate visualization of the aorta, as demonstrated by Sprouse \textit{et al.} \cite{sprouse2004ultrasound}, as well as its non-invasive, safe, and widely accessible nature.

Accurate assessment of the risk of rupture of an aneurysm remains a challenge, although advances in imaging techniques have improved diagnosis and monitoring. Usually, the risk of AAA rupture is assessed by measuring the maximum diameter of the aorta from ultrasound images \cite{brewster2003guidelines}.
However, diameter alone is not a sufficient indicator of rupture risk \cite{khan2015assessing}.
Although diameter analysis has been widely adopted in clinical practice, it does not account for the multi-factorial nature of AAA pathogenesis \cite{nordon2009review}. 
The risk of progression and rupture of an aneurysm is influenced by various factors beyond the geometry, such as the biomechanical properties of the aortic wall \cite{{VORP20071887},{vorp2005biomechanical},{gasser2016biomechanical}}. 
Among these properties, stiffness is particularly significant as it provides direct insight into the structural integrity of the aortic wall \cite{{kadoglou2012arterial},{sonesson1999abdominal}}. Given that ultrasound is routinely used for AAA monitoring, we can go beyond the anatomical assessment of diameter and extract the functional information, such as the vessel stiffness.




To address the inadequacy of diameter alone in AAA risk assessment, an adjunct modality to ultrasound imaging, elastographic strain imaging, has emerged as a promising tool for estimating tissue stiffness \cite{gennisson2013ultrasound}. Elasticity imaging, an additional step to strain imaging, calculates stiffness by solving an inverse problem using elastrography-derived displacement fields measured from ultrasound \cite{sigrist2017ultrasound}. This approach helps to estimate the biomechanical behavior of the aortic vessel, providing a better understanding of aneurysms. Traditionally, iterative techniques have been used to solve the inverse problem. However, these methods often suffer from high computational complexity \cite{richards2011investigating}. Recently, with the rapid advancement of deep learning, data-driven approaches have shown great potential to solve inverse problems more efficiently \cite{tuladhar2023deep}. 


Early work by Fromageau \textit{et al.} \cite{fromageau2005feasibility} demonstrated the feasibility of applying elastography to abdominal aortic aneurysms (AAA) by computing the strain measured from ultrasound data. Similarly, Bonnardeaux \textit{et al.} \cite{bonnardeaux_vivo_2024} tracked the physiologic motion of the aortic wall, to calculate the strain, from the B-mode ultrasound images. Trachet \textit{et al.} \cite{trachet2015performance} conducted a comprehensive {\it in-vivo} study of AAA in mice comparing various ultrasound modalities for strain estimation. These findings also showed that diameter alone is not sufficient to assess the risk of rupture and highlighted the value of computing biomechanical markers. Furthermore, Voizard \textit{et al.} \cite{voizard_feasibility_2020} explored shear wave elastography to calculate stiffness in the aortic sac after endovascular aneurysm repair surgery. They also noted that shear wave signals can go undetected in certain regions, particularly in the posterior wall, limiting consistent quantification.

Several studies have applied ultrasound elastography to AAAs, demonstrating its potential to quantify stiffness variations in aortic vessels. 
Mix \textit{et al.} introduced a pressure-normalized strain imaging approach to detect changes in stiffness. Their method involved measuring strain (the spatial derivative of displacement), which serves as a precursor to estimating the stiffness using inverse elasticity reconstructions \cite{mix2017detecting}. 
Disseldorp \textit{et al.} used 3D ultrasound to track vessel wall displacements and applied an iterative characterization method to estimate stiffness distributions \cite{van2019quantification}.  This approach relied on sophisticated 3D ultrasound probes, which are not commonly available in standard clinical settings. Our method uses 2D ultrasound to capture stiffness variations, offering greater accessibility and clinically useful mechanical insights with simpler imaging hardware.
Long \textit{et al.} proposed a method to quantify the compliance, the inverse of stiffness, from the US Doppler images of AAA \cite{long2005compliance}.

In recent years, deep learning techniques have shown promise in the field of medical imaging and in solving complex problems.
Deep learning can also make complex mappings between the input and output images \cite{isola2017image}.
The use of such a technique has also been extended to the field of ultrasound elastography to calculate the properties of the material \cite{li2022deep}.
Convolutional neural networks (CNNs) in particular have been applied to estimate strain directly from raw radiofrequency (RF) ultrasound data \cite{{wu2018direct},{gao2019learning}}.
To solve the inverse problem and infer stiffness, Tuladhar \textit{et al.} proposed a U-Net-based architecture to reconstruct stiffness maps from US measured displacement fields \cite{tuladhar2025ultrasound}. 

The application of deep learning to AAAs is still an emerging field, but its potential has been highlighted in related work.
Lu et al. \cite{lu2019deepaaa} and Hong et al. \cite{hong2016automatic} employed deep neural networks to segment AAAs in computed tomography (CT) scans, enabling more accurate anatomical localization and measurement. Similiarly, Maas \textit{et al.}  proposed a deep learning approach for automatic AAA segmentation in ultrasound images \cite{maas2024automatic}. However, the integration of deep learning with ultrasound elastography for AAAs remains largely unexplored, leaving a key gap in biomechanical assessment.

To address these gaps, we introduce a deep learning framework for elasticity imaging of abdominal aortic aneurysms. We first generate a comprehensive dataset through finite element (FE) forward modeling, which simulates realistic displacement fields resulting from diverse underlying modulus distributions. We then train a U-net based deep neural network using normalized mean squared error (NMSE) as the loss function to learn the mapping between the displacement field and the corresponding spatial modulus distribution. To validate the model's generalizability and clinical relevance, we test them on various kinds of realistic displacement data measured from simulated, phantom and clinical ultrasound image data. First, we test the trained network on digital phantom data generated via 3D COMSOL simulations. Next, we expand the experimental scope to include physical phantom experiments involving four different types of vessels, designed to mimic varying biomechanical conditions. Finally, we assess the performance of the model on clinical ultrasound exams of human patients with AAA, demonstrating real-world applicability. We evaluated the output of the simulated data using a NMSE comparison of the reconstructed modulus image and the true forward modulus image. In addition, for the simulated data and experimental data, we calculated the ratio of the modulus between regions of the vessel with expected values. We compare these outcome metrics for our deep learning (DL) based predictions with results obtained from an iterative elasticity reconstruction method (ITR), similar to that used in Richards \textit{et. al.} \cite{richards2011investigating}, using the same measured displacement fields.

In summary, the contributions of this study are: (1) the development of a comprehensive and physiologically relevant data set for the imaging of elasticity of aortic vessels; (2) the training of a deep learning model to estimate the distribution of the modulus from the displacement field; (3) the application of this model in simulated, experimental phantom and clinical data, highlighting its potential for clinical translation; and (4) comparison with an iterative method. Unlike prior work, where we focused on strain estimation, our approach here directly targets the reconstruction of elastic modulus fields, thereby advancing the capability of ultrasound elastography to produce quantitative tissue parameter characterization. 

\section{Data generation and acquisition}
\subsection{Simulated Data with Finite Element Modeling}
\label{sec:FEforward}



\begin{figure}[h!]
\centering
\includegraphics[width=1.0\linewidth]{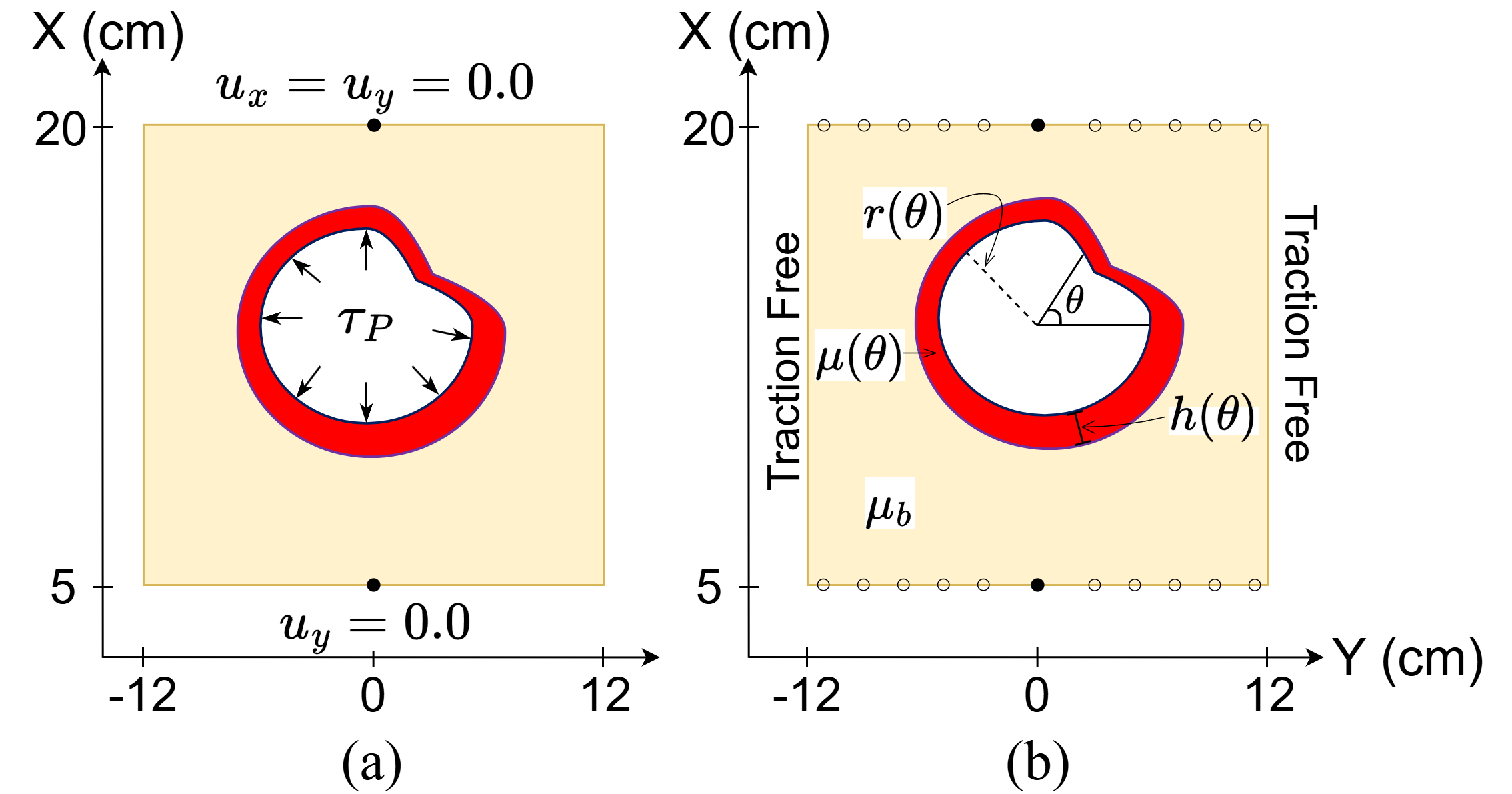}
\caption{FE simulation parameters: (a) illustrates different boundary conditions, and (b) depicts variations in the geometry.}
\label{fig:fem} 
\end{figure}

Our deep learning approach leverages a data-driven framework to model the relationship between displacement fields and modulus distributions. Using a 2D, nearly incompressible plane strain finite element (FE) model, we simulated modulus distributions and their corresponding displacement fields to create a dataset to train the DL model. The FE simulations incorporated the various components as depicted in Figure \ref{fig:fem}. The inputs included the spatial shear modulus distribution representing the 2D modulus distribution of the aortic vessel and the surrounding tissue domain, the vessel's geometry discretized into an FE mesh, and boundary conditions applied at the tissue domain boundary and the inner wall of the vessel. The FE mesh was composed of triangular elements with nodes distributed over a grid spanning [−0.12,0.12] m in the x-direction and [0,0.20] m in the y-direction. The outputs from the model were the displacement fields in the x and y direction, denoted as $u_x(x,y)$ and $u_y(x,y)$, respectively.

To create the spatial modulus distribution, we first constructed an aortic geometry by varying the inner radius of the vessel wall and the wall thickness. We simulated the variability of the inner radius with the following equation:
\begin{equation}
\label{eq:radius}
    r(\theta) = r_o + \sum_{i=1}^{3} A_i cos(2\pi i\theta + \psi).
\end{equation}
Here, $r_o$ is the base radius randomly selected between 1.5 cm and 2.5 cm. Variations in radius were brought about by adding a sum of harmonic perturbations to the base radius, as seen in Equation \ref{eq:radius}. In this equation,  $A_i$ is the amplitude of the $i^{th}$ perturbation and is randomly chosen between 0 and 1 cm. In equation (\ref{eq:radius}), $\psi$ is a random phase with an offset ensuring non-aligned perturbations, and $\theta$ is the angle from 0 to $2\pi$. Thus, the radius varied between 1.5 cm to $\sim4$ cm throughout the dataset which is the typical range for aneurysms \cite{vaitenas2023abdominal}. We also shifted the center of the geometry randomly between $\pm$5 mm in X direction and $\pm$1 mm in the Y direction.

Similarly, for the thickness of the aorta, we use the following equation:
\begin{equation}
    h(\theta) = h_0 + \sum_{i=1}^{2} B_i(1 + cos(2\pi i\theta + \psi)).
\end{equation}
Here, $h_0$ is the base thickness, set as 5 mm, and with similar perturbations applied such that $B_i$ varies between 0 and 5 mm, the vessel thickness is varied between 5 mm to $\sim25$ mm throughout the dataset, which is the estimated range for the AAA patients \cite{koole2013intraluminal}.

We also varied the magnitude of the modulus angularly (around the perimeter of the vessel) with the following equation:
\begin{equation}
\label{eq:modulus}
   \mu_n(\theta) = \mu_0 + \sum_{i=1}^{3} C_i(1 + cos(2\pi i\theta + \psi)),
\end{equation}
where $\mu_0$ is the base shear modulus of 1 kPa and $C_i$ is the amplitude of the $i^{th}$ harmonic variations and is randomly selected between 0.75 and 1.5 kPa. Again, $\psi$ is a random phase offset and $\theta$ is between 0 to $2\pi$. To simulate regions with more drastic changes in mechanical properties, mimicking the heterogeneity in the vessel wall that could arise from localized tissue remodeling, we modified the function of Equation \ref{eq:modulus} as follows:
\begin{equation}
    \mu(\theta)=
\begin{cases}
    \mu_1(\theta) , & \text{if } \theta_0 < \theta < (\theta_0+\theta_1) \\
    \mu_2(\theta), & \text{otherwise}
\end{cases}
\end{equation}
where $\theta_0$ randomly chosen between 0 and $2\pi$ and $\theta_1$ is randomly chosen between $2\pi/3$ and $\pi$. The distribution $\mu_1(\theta)$ was normalized to a range of 1 to 2 kPa, while the distribution $\mu_2(\theta)$ was normalized to a range from 0.25 to 10 kPa. Finally, for half of the generated modulus functions a Gaussian filter, applied angularly, was used to give a smooth transition between regions, thus resulting in modulus distributions with both sharp transitions in simulated material properties and smooth. Ultimately, the values of the shear modulus functions varied between $0.25$ kPa to $10$ kPa throughout vessel regions in the data set. In this work, we only modeled variations angularly around the vessels. Radial variations were not modeled. The shear modulus of the background tissue (outside the vessel) in the models, $\mu_{b}$, was set at a constant of 5 Pa in all simulations.

Variations in the boundary conditions were implemented as follows. The point in the center of the models, as shown in Figure \ref{fig:fem}, was fully fixed in both the x and y directions, that is, $u_x(0.20, 0) = u_y(0.20, 0) = 0.0$. The remaining displacement magnitudes along the top edge (that is, at $x = 0.20$) were randomly selected to be fixed in the x-direction or traction-free, that is, $u_x(0.20,y)=0.0$ or $\tau_x(0.20,y)=0.0$. The remaining boundary conditions on this surface were traction-free (i.e. $\tau_y(0.20,y)=0.0$). Similarly, the bottom point, as shown in Figure \ref{fig:fem}, of all models was fixed in the y direction, i.e. $u_y(0.05, 0)=0.0$ and the remaining bottom edge was randomly chosen to be fixed in the x direction (i.e. $u_x(0.05,y)=0.0$) or traction-free ($\tau_x(0.50,y)=0.0$). All other outer boundary conditions were traction-free. We used a traction boundary condition on the vessel inner lumen, approximating a hydrostatic pressure normal to the surface of the lumen as follows:
\begin{equation}
    \tau_P = P\boldsymbol{n}, \label{eq:tau}
\end{equation}
where, $\boldsymbol{n}$ is the unit vector function normal to the inner lumen wall (see Figure \ref{fig:fem}(a)). The magnitude of the pressure ($P$) was also varied randomly between 1 to 8 kPa.

In this way, by systematically varying the geometry of the vessel, the modulus distributions, and the applied boundary conditions, we generated a comprehensive dataset consisting of paired samples of the spatial modulus distribution and their corresponding displacement fields. To further enhance the variability, we also added random translations to the displacement fields. Furthermore, prior to training our DL model, we normalized the displacement fields by the applied pressure, resulting in pressure-normalized displacement fields that allow quantitative modulus estimation \cite{mix2017detecting}:
\begin{align}
    \tilde{u}^x(\bs{x}) &= u_x(\bs{x})/P, \text{and}\\
    \tilde{u}^y(\bs{x}) &= u_y(\bs{x})/P. \label{eq:normP}
\end{align}
Lastly, the displacement fields and the modulus fields were interpolated from the FE mesh to a regularly spaced grid, which was the same for all models, of 128×128 pixels, with a pixel spacing in $x$ and $y$ of 0.86 mm. (see Figure \ref{fig:phantom_mesh}(c)). The pixels in the interpolated images that were located outside the vessel FE mesh elements were set to 0. This provided our deep learning model with an implicit segmentation of the vessel lumen. This data set, consisting of input images $\tilde{u}^x_{ij}$ and $\tilde{u}^y_{ij}$ and output image $\mu_{ij}$, forms the basis for training our deep learning model to solve the inverse problem in elastography. Our training set includes 30,000 paired examples of 128×128 sized displacement and modulus images. Similarly, we created a validation and test dataset containing 3,000 examples each for a total of 36,000 forward, FE simulated models.



\subsection{COMSOL Data}
A central, transverse slice of displacement data was acquired from 3D FE simulations (COMSOL) with known modulus variations of the ventral wall of the aorta to create simulated US images of the deforming aorta using a point spread function approximating a curvilinear transducer.

\begin{figure}[h]
    \centering
    \includegraphics[width=1.0\linewidth]{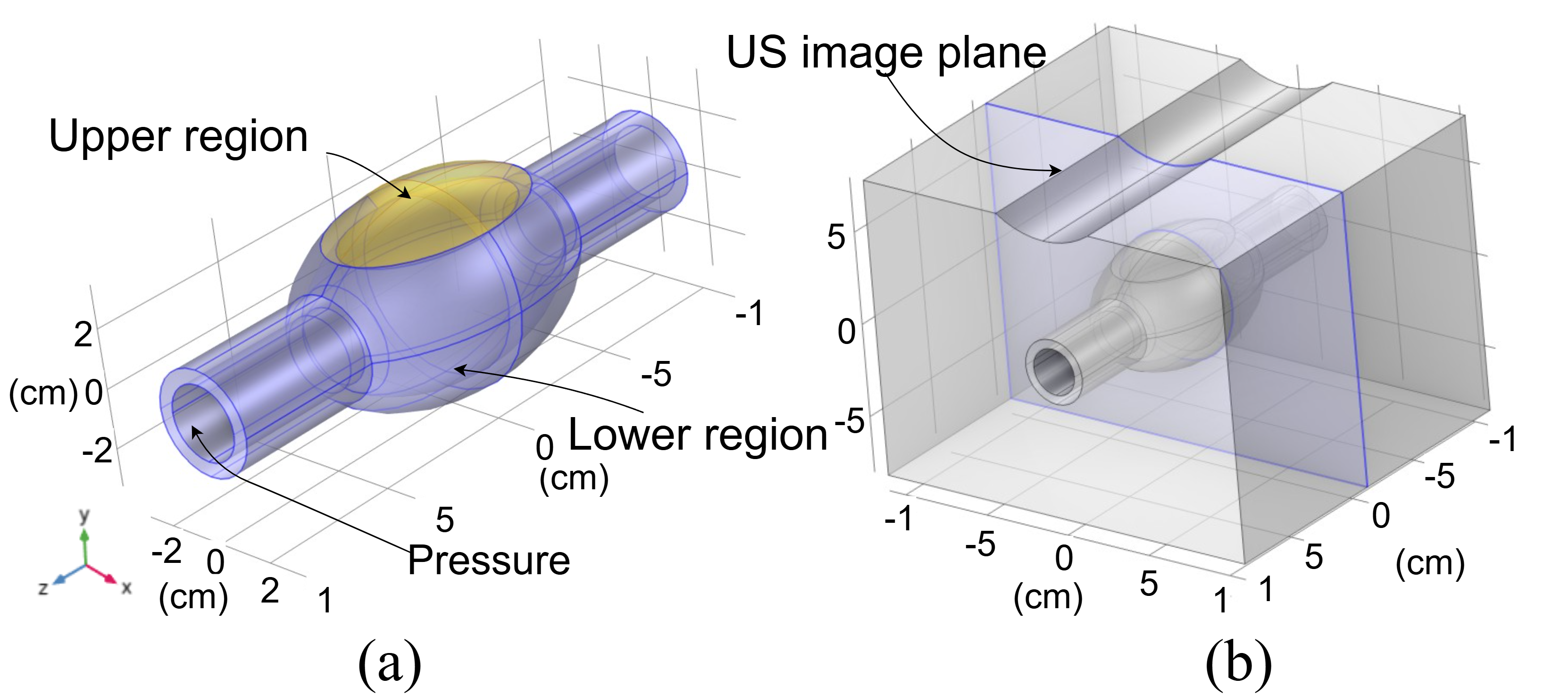}
    \caption{COMSOL 3D simulation model. (a) highlights the upper region where the modulus is varied and the lower region where it remains constant, and (b) shows the extracted US image plane.}
    \label{fig:comsim3d}
\end{figure}

A 3D finite element (FE) model of the abdominal aorta was constructed in COMSOL Multiphysics software to simulate physiologically relevant deformation under a pressure load. 
The model geometry consisted of a cylindrical segment with a localized aneurysmal bulge, representing the aneursym. The aortic wall was modeled as a linearly elastic, nearly incompressible material (with poisson's ratio of 0.495) with regionally varying Young's modulus. Material properties were specified by dividing the wall into two longitudinal zones as shown in Figure \ref{fig:comsim3d}, with the lower region assigned a fixed Young’s modulus of 200 kPa and the upper region assigned a varying modulus within values of 100, 200, 400, and 800 kPa to simulate stiffness contrasts. The Young's modulus of the background tissue, as shown in gray in Figure \ref{fig:comsim3d}(b), was set to 10 kPa in all models. The FE mesh was generated using tetrahedral elements.  The average size of the mesh was 3.25 $mm^3$ inside the vessel and 23.6 $mm^3$ outside the vessel. A static internal pressure ($P_{CM}$) of 5.33 kPa, mimicking pulse pressure of 40 mmHg, was applied to the inner lumen. The outer anterior surface, where the ultrasound transducer would contact the tissue, was fixed to mimic the transducer-tissue interaction. 
The linear elastic solver computed the full 3D displacement field across the aortic vessel. From this output, a central, transverse 2D slice orthogonal to the vessel's axis was extracted, capturing the displacement behavior across the aneurysmal region. 
This 2D displacement field was used to simulate realistic radio-frequency (RF) US images by convolution of a 2-D point spread function (PSF) with point scatterers in an image plane. The PSF was defined to mimic the point response of the Ultrasonix C7-3/50 convex transducer \cite{mix2017detecting}. RF US of deformed tissue were created by deforming the same point scatterers using COMSOL generated displacement fields, within the plane of the image, and repeating the image simulation process.

\subsection{Phantom Data}
Aorta mimicking US phantoms were also created, with varying material properties, and imaged with a 5 MHz, curvilinear US transducer \cite{mix2017detecting}. Displacements were tracked and accumulated from diastolic to systolic pressure via registration of the RF image sequences and a novel regularization.

Phantom data were acquired using an experimental setup as described in Mix \textit{et al.} \cite{mix2017detecting}. The phantoms, previously developed and characterized as shown in Table \ref{tab:phantomexpected}, were submerged in a water bath and connected to a hemodynamic pump designed to produce physiologic cyclic blood flow profiles. The pressure profile, measured by a catheter-based pressure sensor (Millar Mikro-Cath, Auckland, New Zealand) proximal to the ultrasound imaging plane, was synchronized with ultrasound frames using a National Instruments USB-6251 DAQ (National Instruments, Austin, TX, USA). The pulse pressure for these phantoms ($P_{ph}$) was calculated as the difference between the maximum and minimum pressures measured via the catheter during image acquisition. Ultrasound imaging was performed with a Sonix-TOUCH system (Analogic) and a C7–3/50 convex transducer at a center frequency of 5 MHz. Imaging settings included a 60\% sector, 12 cm depth, and a single focus at the posterior wall. RF data were collected at a frame rate of 72 Hz over a 10-second interval (720 frames), corresponding to 10 cardiac cycles.

\begin{figure}
    \centering
    \includegraphics[width=0.99\linewidth]{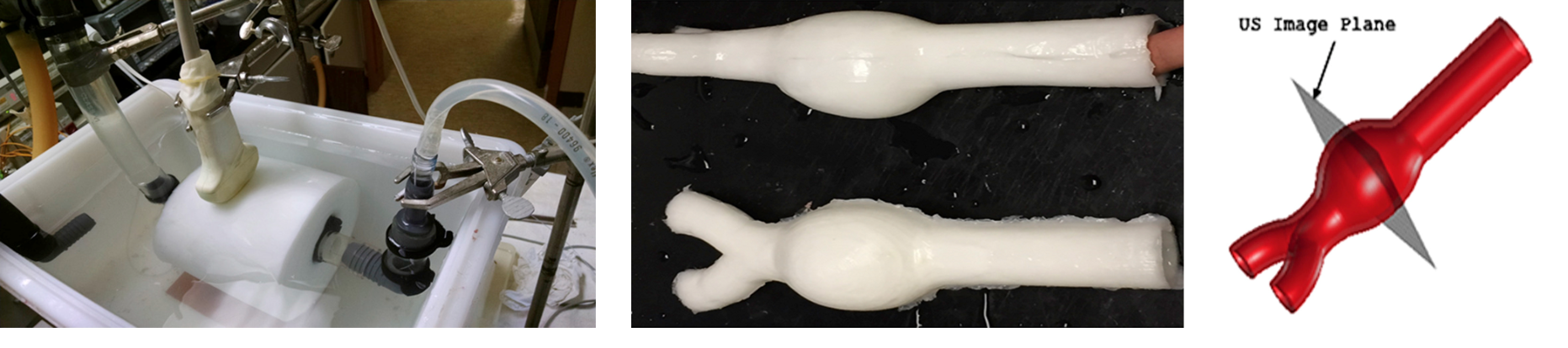}
    \caption{Experimental phantom setup and phantom model \cite{mix2017detecting}}
    \label{fig:enter-label}
\end{figure}

\begin{table}[h]

\centering
\caption{Material properties of the anterior and posterior ends of the phantoms.}
\label{tab:phantomexpected}
\begin{tabular}{c|c|c|c|c}
\hline
\textbf{Phantom} & \makecell{\textbf{Anterior End}\\\textbf{Material}\\\ \textbf{(PVA \% wt)}} & \boldmath{$\mu$}\textbf{(kPa)} & \makecell{\textbf{Posterior End}\\\textbf{Material}\\\ \textbf{(PVA \% wt)}} & \boldmath{$\mu$}\textbf{(kPa)} \\
\hline
1 & 10\% &  $17.4 \!\pm\! 1$  & 10\%  &  $17.4 \!\pm\! 1$  \\
2 & 15\% & $48.3 \!\pm\! 5.7$  & 10\%  & $17.4 \!\pm\! 1$  \\
3 & 20\% & $95.1 \!\pm\! 0.4$  & 10\%  &  $17.4 \!\pm\! 1$  \\
4 & 25\% & $170 \!\pm\! 4.1$ & 10\%  &  $17.4 \!\pm\! 1$  \\
\hline
\end{tabular}
\end{table}

\subsection{Clinical Data}

Patients were prospectively recruited from the vascular surgery department at the University of Rochester Medical Center between July 2015 and August 2016. Eligibility was based on clinical need for a standard ultrasound screening exam for a suspected, newly diagnosed, or previously identified non-repaired abdominal aortic aneurysm (infrarenal or suprarenal). Exclusion criteria included individuals unable to provide informed consent, minors, and members of protected populations such as pregnant individuals and prisoners. Written informed consent was obtained from all participants prior to imaging. For each enrolled patient, raw RF image data was collected using the same ultrasound machine and imaging settings as those used for phantom acquisitions. Systolic and diastolic frames were identified, via visual inspection, using B-mode image sequences. In addition, blood pressure measurements were taken immediately before and after ultrasound examination using a pressure cuff. The patient's pulse pressure ($P_{cl}$) was then calculated as the difference between the recorded systolic and diastolic blood pressures, converted to kPa, averaged between the two measurements. \cite{mix2017detecting}. Further details of the clinical protocol can be found in Zottola \textit{et al.} \cite{zottola2023intermediate}.

\section{Image processing and analysis}
\subsection{Image registration for displacement estimation}
\label{sec:ir}

An image registration algorithm was used to measure tissue displacements from RF US images for phantom and clinical data. 
We calculated the total accumulated displacement field within the vessel wall and surrounding tissue (excluding the blood lumen) from diastole to systole to capture the vessel wall deformation. The details of the registration algorithm can be found in Tuladhar \textit{et al.} \cite{tuladhar2025ultrasound}. Briefly, displacements were computed sequentially, frame-to-frame, and accumulated with respect to the diastolic frame as a reference. The optimization functional for the {\textit i\textsuperscript{th}} frame-to-frame displacement is:
\begin{align}
\pi[\bu_i(\bx)] &= \frac{1}{2} \int_{\Omega} \left(I_{i\text{-}1}(\bx+\bu_{i\text{-}1}(\bx)) - I_{i}(\bx + \bu_{i}(\bx)) \right)^2 \ d\Omega \nonumber \\
&\quad + \alpha \int_{\Omega} \|\nabla \cdot \bs{A}_i(\bx)\| \ d\Omega,
\label{eq:mreg}
\end{align}
where $\bs{A}_i(\bx)$ is the constitutive relation for a linear elastic, homogeneous, incompressible material deforming under plane strain and can be written as:
\beq
\bs{A}_i(\bx) = \frac{2\nu}{1-2\nu}  (\nabla \cdot \bu_i(\bx)) \I+ (\nabla \bu_i(\bx) + \nabla \bu_i(\bx)^T). 
\label{eq:const}
\eeq

In Equation \ref{eq:mreg} the first term estimates the displacement field that best captures the deformation of the tissue, as realized by spatial shifts of image intensity, while the second term is a regularization that penalizes deformation patterns that are physically inconsistent with plain strain. In this work, a value of $\nu=0.45$ was used, as the tissue was assumed to be nearly incompressible. The frames corresponding to minimum diastole, where $\bu_0(\bx)$$=$$\mathbf{0}$, and maximum systole were manually identified by a vascular surgeon based on B-mode image sequences \cite{mix2017detecting}. It should be noted that, unlike in Mix et. al. \cite{mix2017detecting}, no accumulation of displacement fields was performed here. The displacements were measured from only two frames that corresponded to the point of minimum diastolic pressure and the maximum systolic pressure. Segmentation of the boundary between the vessel and blood and the boundary between the vessel and surrounding tissue was manually segmented by a vascular surgeon using the B-mode image at minimum diastole \cite{mix2017detecting}. Figure \ref{fig:phantom_mesh}(a) shows an example B-mode image of a phantom. Displacement fields were measured within the vessel wall and surrounding tissue on a finite element (FE) discretized mesh of quadrilateral elements, created using open-source mesh generation software \cite{GeuzaineRemacle2009}. The average area of the FE elements were 0.027 $\pm$ 0.015 cm$^2$. An example FE mesh is shown in Figure \ref{fig:phantom_mesh}(b).

The output of the image registration algorithm was the axial and lateral components of the displacement field, within the vessel wall and surrounding tissue on the FE mesh. For all displacement field measurements to be input into the deep learning model, the displacement field grid values within the vessel were interpolated onto a regular grid as shown in Figure \ref{fig:phantom_mesh}(c). Again, the gridded values within the surrounding tissue and blood regions were zeroed for both components. Normalized displacement fields, $\tilde{u}^x_{ij}$ and $\tilde{u}^y_{ij}$, were found by dividing the COMSOL measurements with the forcing pressure BC used in the forward simulation ($P_{CM}=5.33$ kPa), the phantom measurements by the pulse pressure measured independently during the experiments ($P_{ph}$) and the clinical measurements by the measured pulse pressure ($P_{cl}$) found using the blood pressure cuff at the time of clinical scan.  

The input to the iterative reconstruction algorithm was the displacement field on the registration mesh, both within the vessel and within the surrounding tissue, and thus no interpolation or normalization was performed for those input data.

\subsection{Iterative method for modulus reconstruction}

A model-based optimization algorithm was used to reconstruct the modulus distribution ($\mu_{itr}(\bx)$) from the measured displacement fields of the COMSOL, phantom, and clinically acquired US image data. The algorithm is based on a soft prior reconstruction technique described in Richards \textit{et al.} \cite{richards2011investigating}, previously developed for intravascular ultrasound elastography. First, a 2D finite element model was created, using a constitutive equation for a nearly incompressible material that deforms in plane strain, with two domains, the vessel and the surrounding tissue, as shown in Figure \ref{fig:phantom_mesh}. The Poisson ratio of the entire domain was set to $\nu=0.45$ and the shear modulus values within the vessel domain are the optimized parameters. A 2D traction vector function, $\tau_{o}$, is applied to the outermost edge of the domain of this model and a pressure boundary condition, $\tau_{P}$ with magnitude $P_{it}$ such that $\tau_{P}(\bs{x})=P_{it}\,\bs{n}(\bs{x})$, is applied to the blood lumen wall. Here, $\bs{n}(\bs{x})$ is a unit vector function defined on the inner lumen, $\Gamma_P$, that is everywhere normal to that boundary.

The iterative algorithm used here consists of two steps. The first step solves for the boundary conditions of the desired domain, keeping the modulus distribution fixed, which minimizes the difference between the total measured displacement field ($\bu_m(\bx)$), from diastole to max systole, and a model-predicted displacement field ($\bu_p(\bx)$), in the least squares sense. The optimization functional for this step is as follows:
\begin{equation}
\pi_B[\tau_o,P_{it}] = \frac{1}{2} \int_{\Omega} \left\| \bu_p[\tau_o,P_{it}] -  \bu_m \right\|^2 d\Omega,\label{eq:FEo}
\end{equation}
where $\Omega$ is the modeled domain. Using the same FE discretization as was used in the registration algorithm described in Section \ref{sec:ir}, we can rewrite Equation \ref{eq:FEo} as the following matrix equation:
\begin{equation}
\overline{\pi}_B = \left( \overline{u}_p - \overline{u}_m \right)^T D \left( \overline{u}_p - \overline{u}_m \right ),\label{eq:FEod}
\end{equation}
where $\overline{u}_p$ is an array of both components ($x$ and $y$) of the discrete displacement fields, defined at the nodal positions, and $\overline{u}_m$ is an array of the same size consisting of the measured values of the displacement fields. Here, $D$ is a matrix that performs the integration of shape functions for the least squares difference of the measured and predicted displacement fields using a 3$\times$3 Gauss quadrature integration.

To constrain the displacement fields in Equation \ref{eq:FEo}, we also define a forward FE model for a nearly incompressible material deforming in 2D plane deformation (within $\Omega$) with the following weak form of a static linear elastic material (neglecting body forces) as:
\begin{equation}
\int_{\Omega} \left ( \nabla(G(\bx) \bs{A}_p(\bx) \right ) \cdot \bs{w}(\bx) \ d\Omega = \bs{0} \label{eq:FE1}
\end{equation}
where $\bs{A}_p(\bx)$ is defined in Equation \ref{eq:const} with $\nu=0.45$, $G(\bs{x})$ is a scaled version of the modulus distribution, and $\bs{w}(\bs{x})$ are the weighting functions of the displacement field \cite{hughes2003finite}. In this algorithm, we chose to use a unitless modulus function, $G(\bs{x})$, where the regularization keeps the mean of $G(\bs{x})$ approximately equal to 1 within the vessel domain. We then search for the boundary conditions that result in the best match to our measured displacement field, given that modulus. The modulus is then scaled to a quantitative magnitude in post-processing. With only traction boundary conditions, integration by parts can be used to rearrange Equation \ref{eq:FE1} as (dropping the explicit dependence on $\bx$):
\begin{equation}
\int_{\Omega}  \nabla \bs{w} \cdot \left(G \ \bs{A}_p \right) \ d\Omega  = \int_{\Gamma} \bs{w} \cdot \bs{t} \ d\Gamma. \label{eq:FE2}
\end{equation}
Here, $\bs{t}$ are the traction boundary conditions that we define to exist on the entire boundary of the domain, $\Gamma$. The boundary can also be divided into the inner lumen boundary $\Gamma_P$ and the outer boundary of the background tissue $\Gamma_o$ such that $\Gamma = \Gamma_P \cup \Gamma_o$. We then define the traction boundary conditions in 3 parts: the pressure in the inner lumen, $P_{it}$, the traction vector function on the outer boundary, $\tau_o$, and a weak spring-like traction on the outer boundary, necessary to avoid rigid body translations and rotations and to keep the predicted displacements close to the measured values on that boundary. Thus, Equation \ref{eq:FE2} can be written more explicitly as:
\begin{multline}
    \int_{\Omega}  \nabla \bs{w} \cdot \left(G \ \bs{A}_p\right) \ d\Omega  = \int_{\Gamma_{o}} \bs{w} \cdot \tau_o \ d\Gamma_{o} +...\\ k_s \int_{\Gamma_o} \bs{w} \cdot (\bu_m-\bu_p) \ d\Gamma_{o} +\int_{\Gamma_{P}} \bs{w} \cdot {\tau_P} \ d\Gamma_{P}. \label{eq:FE3}
\end{multline}
Here $\bs{n}$ is a function defining the normal vector to the inner lumen wall as shown in Figure \ref{fig:fem}(a). Assuming that $\mu(\bx)$ is known and fixed and using the same FE discretization as above, we can plug Equation \ref{eq:tau} into Equation \ref{eq:FE3} and perform the integrations to find the following matrix equation:
\begin{equation}
K \overline{u}_p = F_o \overline{\tau}_o + k_s F_o (\overline{u}_m-\overline{u}_p) + f_P P_{it}, \label{eq:FE4}
\end{equation}
where $K$ is the stiffness matrix representing the integrations of the LHS of Equation \ref{eq:FE3}, performed within quadrilateral elements using a 3 by 3 Gauss quadrature numerical integration with reduced integration of the displacement dilatation term of $\bs{A}_p(\bx)$ to avoid mesh locking \cite{hughes2003finite}. Then, $\overline{\tau}_o$ is an array of both components of the discrete traction on the outer boundary, only nonzero at the nodes on the outer boundary. $F_o$ is a matrix that performs the integration of the first term of the RHS of Equation \ref{eq:FE3} using a 1D, 3 point Gauss integration method on the edges of the elements along the outer surface of the domain. This matrix is the same size as $K$ and mostly sparse, except for the row/column locations corresponding to nodes on the outer boundary. Lastly, $f_P$ is an array the same size as $\overline{\tau}_o^T$ that corresponds to the integration of $\int_{\Gamma_{P}} \bs{w} \cdot \bs{n} \ d\Gamma_{P}$, again using a 1D, 3 point Gaussian integration method. Equation \ref{eq:FE4} can be rearranged to solve for $\overline{u}_p$ as follows:
\begin{equation}
\label{eq:usolv}
\overline{u}_p = (K+k_s F_o)^{-1} [ F_o \overline{\tau}_o + f_P P + k_s F_o \overline{u}_m].
\end{equation}
and plugged into Equation \ref{eq:FEod} to result in a linear equation for the boundary parameters (that is, $\tau_o$ and $P$), which can be solved directly via linear least squares. In all inversions done in this work, $k_s=1e$-$3$ was chosen empirically such that $k_s(\bu_m -  \bu_p) \ll \tau_o$ but large enough to keep the matrix inversion in Equation \ref{eq:usolv} well posed.

The second step of the reconstruction method solves for the modulus distribution ($G(\bx)$) of the desired domain, keeping the boundary conditions found in the first step, $\tau_o$ and $P$, fixed. Here, the difference between the total measured displacement field and a model-predicted displacement field is minimized, as in the first step, in the least squares sense. However, the modulus field in the second step is updated iteratively using a quasi-Newton method. The details of the optimization algorithm were published previously in Richards \textit{et al.} \cite{richards2011investigating}. The algorithm developed there was adapted for use in this paper. As in Richards \textit{et al.} \cite{richards2011investigating}, the second step also uses \textit{a priori} information for modulus recovery, incorporated into the regularization, to improve the ill-posedness of the inverse problem. The optimization functional utilized here is as follows:
\begin{align}
\pi_\mu[\hat{G}(\bx)] &= \frac{1}{2} \int_{\Omega} \left\| \bu_p[G(\bx)] -  \bu_m \right\|^2 d\Omega \nonumber \\
&+ \alpha \int_{\Omega} \left\| \nabla c(G(\bx),G_0(\bx)) \right\| d\Omega. \label{muupdate}
\end{align}
Here, $G(\bx)$ is the modulus distribution within the entire solid tissue domain $\Omega$, $\hat{G}(\bx)$ are the modulus values that fall within the wall of the vessel and $G_0(\bx)$ is the nominal function that serves as the prior spatial information to which the modulus is regularized. In this work, the value of $G_0(\bx)$ within the vessel was set to a value of 1 and the value in the surrounding tissue was set to 0.1. Thus, we make the implicit assumption that the segmented surrounding tissue has a shear modulus that is 1 order of magnitude lower than the logarithmic mean of the modulus within the vessel. These modulus values also served as the initial guess to the modulus field. Note that we only search for, and thus update, the values of the modulus within the vessel. Thus, the modulus values within the surrounding tissue remain constant at 0.1. The function $c(G,G_0)$ here serves as the logarithmic regularization function, used to remove the dependence of the regularization on the general magnitude of the modulus and the assumed contrast between the modulus of the vessel and the surrounding tissue:
\begin{equation}
c(G,G_0)=ln(G/G_0).
\end{equation}
Thus, the regularization used here is a TVD-type regularization of the logarithmic function $c(G,G_0)$ \cite{richards2011investigating}. Only the modulus values within the vessel (that is, $\hat{G}(\bx)$) were allowed to update. The modulus of the background tissue was held fixed. Equation \ref{eq:FE3} was used, with $\tau_o$ and $P$ fixed, as the constraint equation of the model and an adjoint method was used to calculate the gradient of Equation \ref{muupdate} as described by Richards \textit{et al.} \cite{richards2011investigating}. Optimization was performed using Broyden–Fletcher–Goldfarb–Shanno (BFGS) algorithm  with 5 update iterations each time the second step was performed.

All reconstruction methods were implemented in MATLAB (The Mathworks Inc., Natick, MA, United States). The stiffness matrix and boundary integration matrices were stored as sparse matrices in MATLAB. We ran 30 total iterations (Step 1 then Step 2) to arrive at the output reconstructed modulus. Furthermore, we scaled the reconstructed modulus field with the independently measured pulse pressure to obtain the quantitative value of the modulus in kPa (i.e., $\mu_{itr}(\bx) = PP \times (G(\bx)/P_{it})$), where $PP$ was equal to $P_{CM}$, $P_{ph}$, or $P_{cl}$ depending on the data source, and interpolated from the FE mesh onto the same uniform spatial grid of 128$\times$128 pixels, as described in Section \ref{sec:FEforward}, to arrive at our discrete, quantitative modulus image, $\mu_{ij}^{itr}$.

\subsection{Deep learning for modulus reconstruction}

\label{sec:dl}


\begin{figure}[h]
\begin{center}
\begin{tabular}{c} 
\includegraphics[height=3.6cm]{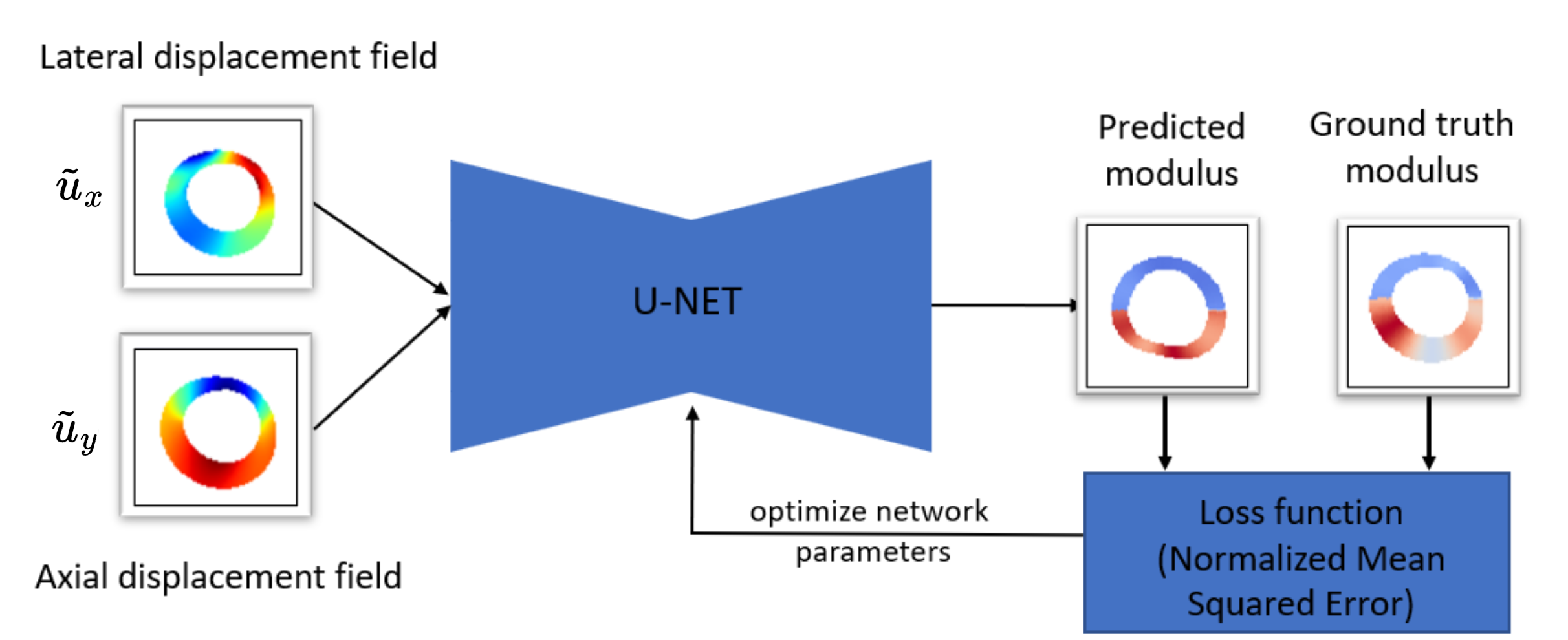}
\end{tabular}
\end{center}
\caption{Deep learning U-NET model featuring a NMSE loss function for optimization.}
\label{fig3:blockdiagram}
\end{figure} 

\subsubsection{Deep learning framework}


We used a deep learning framework of convolutional neural networks (CNNs) to learn the mapping between the pressure-normalized displacement fields and the spatial distribution of modulus as shown schematically in Figure \ref{fig3:blockdiagram}. First, the computational domain was discretized into a uniformly spaced grid of $128 \times 128$ pixels. Each point in the grid $(i,j)$ corresponded to a spatial location, where $\tilde{u}_{ij}^x$ and $\tilde{u}_{ij}^y$ denote the axial and lateral components of the input displacement fields, respectively, and $\mu_{ij}$ represents the modulus output image.  As discussed in Section \ref{sec:FEforward}, only those values of the input and output images that were within the vessel wall had non-zero values. This served to provide our CNNs with an implicit vessel geometry. To ensure a consistent scale for the DL model and stabilize training \cite{sola1997importance}, we further normalize our input displacement fields
 by the maximum of displacement magnitude, $n_m$, calculated as:
\begin{equation}
n_m = \max_{\ i,j } \sqrt{{(\tilde{u}_{ij}^x})^2+{(\tilde{u}_{ij}^y})^2},
\end{equation}
where $()^2$ denotes a pixel-wise squaring. Then, we divide the displacement field images:
\begin{equation} 
\hat{u}_{ij}^x  = \tilde{u}_{ij}^x / n_m; \, \hat{u}_{ij}^y  = \tilde{u}_{ij}^y / n_m,
\end{equation}
and normalize the corresponding modulus image by multiplying with the same factor:
\begin{equation}
G_{ij}  = \mu_{ij}  \times n_m.
\end{equation}
This results in normalized axial ($\hat{u}_{ij}^x$) and lateral ($\hat{u}_{ij}^y$) displacement field images and a normalized modulus image, $G_{ij}$, used to train our DL model. All output modulus images output from the fully trained model are rescaled back to a quantitative modulus such that the values are in units of kPa:
\begin{equation}
\label{eq:rescalemod}
 \mu_{ij}^{dl}   = G_{ij} / n_m.
\end{equation}

This normalization and rescaling procedure was applied consistently for both training and inference.

\subsubsection{Network Architecture}

We utilized a U-net based architecture \cite{ronneberger2015u}, which is well-suited for tasks requiring precise localization and spatial understanding. Additionally, we drew inspiration from the Pix-to-Pix framework \cite{isola2017image} widely used for image-to-image translation. However, in our implementation, we used the generator component with a loss function as it was more appropriate with our objective of predicting continuous-valued outputs.  This neural network comprised of 7 downsampling layers in the encoder and 7 upsampling layers in the decoder, with a latent dimension of 512. 

\subsubsection{Loss function}

To train the model, we employed the normalized mean squared error (NMSE) as the loss function between the ground truth normalized modulus ($G^t_{ij} $) and the predicted normalized modulus ($G^p_{ij} $), defined as:


\begin{equation}
\label{eq:nmse}
\mathcal{L}_{\text{NMSE}} = \text{NMSE}_{} \ \!\langle G^t_{ij} , G^p_{ij}  \ \!\rangle,
\end{equation}
where,
\begin{equation}
\label{eq:nmse2}
\text{NMSE}\ \!\langle A_{ij},B_{ij} \ \!\rangle := 
\frac{\sum\limits_{i=1}^{128} \sum\limits_{j=1}^{128} (A_{ij} - B_{ij})^2}
     {\sum\limits_{i=1}^{128} \sum\limits_{j=1}^{128} (A_{ij})^2}, 
\end{equation}
and the $()^2$ operations are applied pixel-wise.

\subsubsection{Evaluation Metrics}

To evaluate the performance of the DL model on the validation and test sets, we calculated the NMSE loss between the predicted rescaled modulus $ \mu_{ij}^{dl} $, obtained from Equation \ref{eq:rescalemod}, and the corresponding ground truth modulus ($\mu_{ij} $) from the dataset, calculated using Equation \ref{eq:nmse2}, (i.e., $\text{NMSE}\ \!\langle \mu_{ij},\mu_{ij}^{dl} \ \!\rangle$).

In addition, to assess the model’s ability to accurately localize the aorta, we calculate the dice score coefficient (DSC) \cite{dice1945measures}. To calculate DSC, we first created binary masks ($B_{ij}^t $) from ground truth modulus distribution by:


\begin{equation}
    B_{ij}^t  = 
    \begin{cases}
         1,   \text{   if } \mu_{ij} >0 \\
        0, \text{  otherwise}.
    \end{cases}
\end{equation}

The binary mask from the predicted modulus distribution ($B_{ij}^p $) was computed similarly. The overlap between $B_{ij}^t $ and $B_{ij}^p $ was then measured using the DSC:

\begin{equation}
\textit{DSC} = 
\frac{2 \sum_{i=1}^{128} \sum_{j=1}^{128} \left( B_{ij}^t \odot B_{ij}^p \right)}
     {\sum_{i=1}^{128} \sum_{j=1}^{128} B_{ij}^t \; + \; \sum_{i=1}^{128} \sum_{j=1}^{128} B_{ij}^p },
\end{equation}
where $\odot$ is the element-wise Hadamard product. The DSC score ranges from 0 to 1, where 1 indicates perfect overlap between the predicted and true mask, and 0 indicates no overlap.  


\subsubsection{Training and validation}

\begin{figure}
    \centering
    \includegraphics[width=1.0\linewidth]{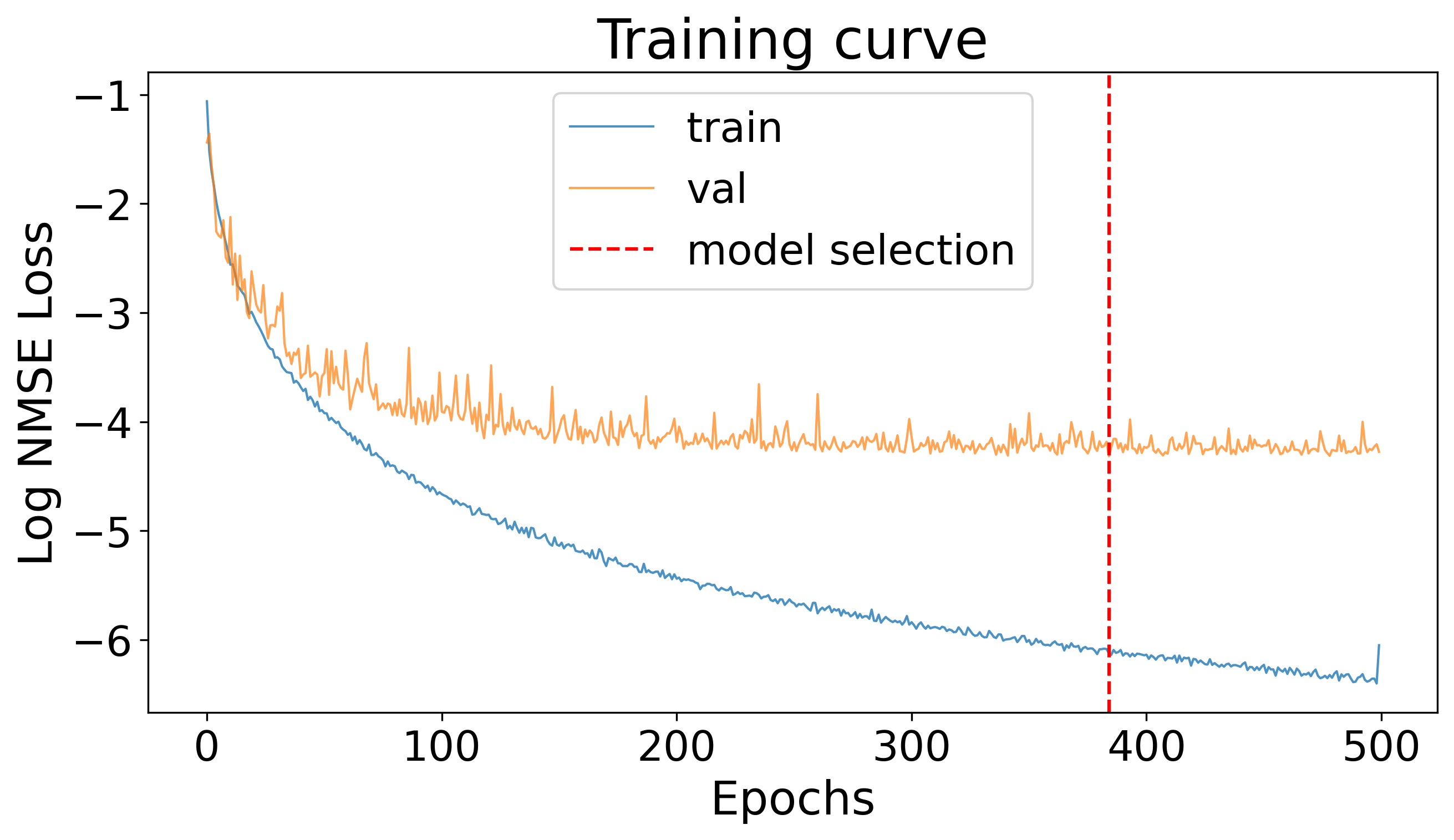}
    \caption{Training curve for the deep learning model}
    \label{fig:curve}
\end{figure}

During training, model performance was monitored using the validation set, evaluating the performance at regular intervals. The model checkpoint corresponding to the lowest validation NMSE was selected for further evaluation. We trained our DL model for 500 epochs. Figure \ref{fig:curve} shows the NMSE loss for both the training set (blue) and the validation set (orange) across the training process. The final model was selected at the epoch corresponding to the minimum validation loss, which occurred at epoch 384 with an NMSE of 0.013. After training completion, the model was applied to the displacement data from four domains — simulated test cases, COMSOL-generated digital phantoms, tissue-mimicking phantom experiments, and clinical datasets, to assess generalizability and robustness.

\section{Results}
\subsection{Simulated Test Data}

\begin{figure}[h]
    \centering
    \includegraphics[width=0.99\linewidth]{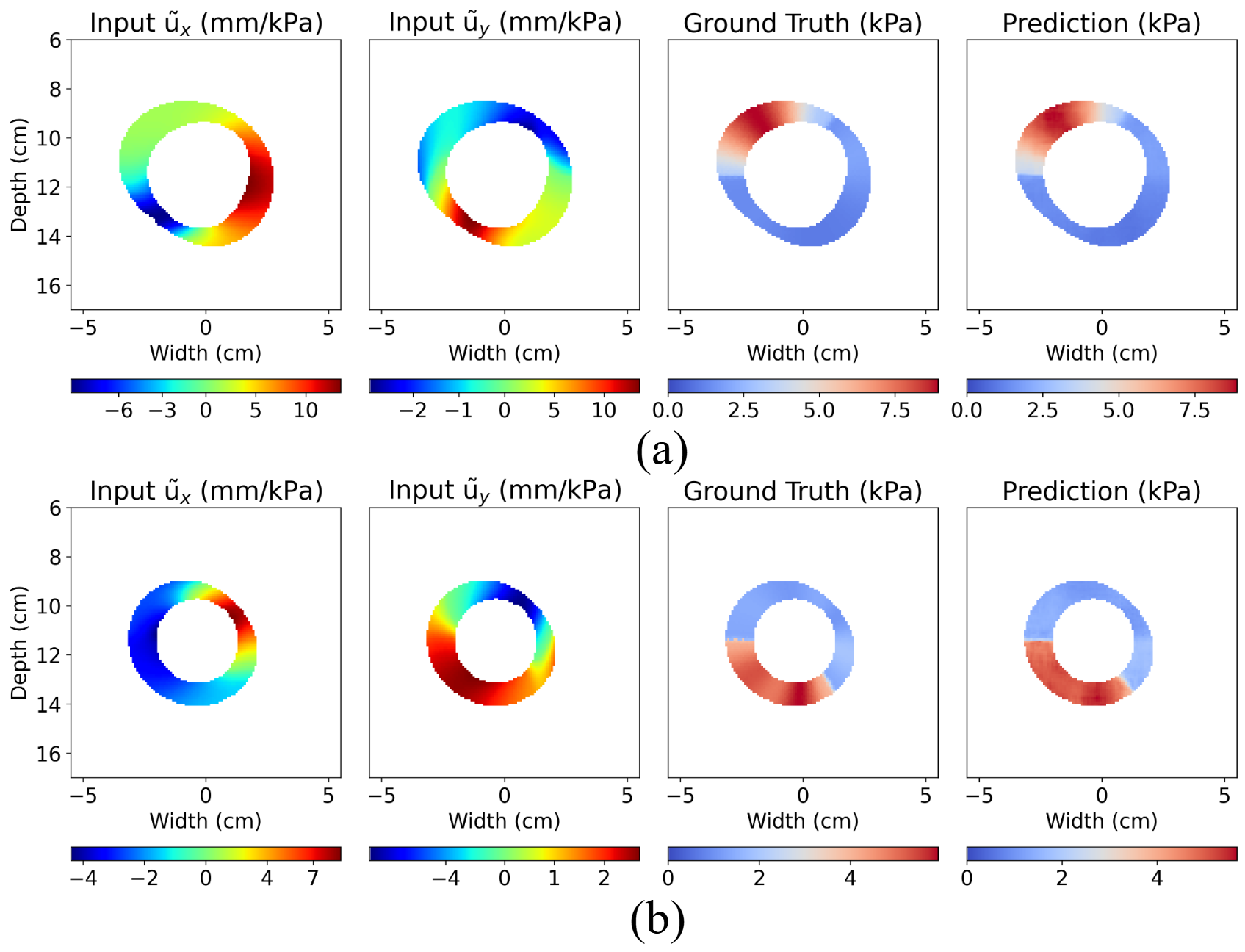}
    \caption{Simulated test examples showing pressure-normalized X and Y displacement fields as inputs to the DL model, along with the corresponding ground truth and predicted modulus distributions.}
    \label{fig:sim_all}
\end{figure}

We evaluated our DL model on the test dataset to assess its performance. Our DL model achieved a mean NMSE error of 0.0073 across the test dataset.  A detailed summary of various performance metrics is presented in Table \ref{tab:results_sim}. We also present two samples from the test set in Figure \ref{fig:sim_all}. The figure shows the two input diplacement fields, the ground truth and the predicted modulus distribution. For example (a), the NMSE is 0.0037 and the DSC is 1.0, and similarly for example (b), the NMSE is 0.0014 and the DSC is 1.0.

\begin{table}[h]
\captionsetup{font=small}
\caption{Summary of mean of various evaluation metics for the simulated test dataset} 
\label{tab:results_sim}
\begin{center}       
\small
\begin{tabular}{|c|c|} 
\hline
\rule[-1ex]{0pt}{3.5ex}  
\textbf{No. of test images}  & $3,000$ \\
\hline
\textbf{Mean NMSE} & $0.0073$ \\
\hline
\textbf{Mean DSC} & $0.999$ \\
\hline
\end{tabular}
\end{center}
\end{table}

These results demonstrate that the model performs well on the simulated data. The low NMSE values and high DSC score confirm that the predicted outputs closely match the ground truth, both quantitatively and spatially. We further extend the model to the real-world data.

\subsection{Comsol Data}
\begin{figure}[h]
\centering
\includegraphics[width=0.99\linewidth]{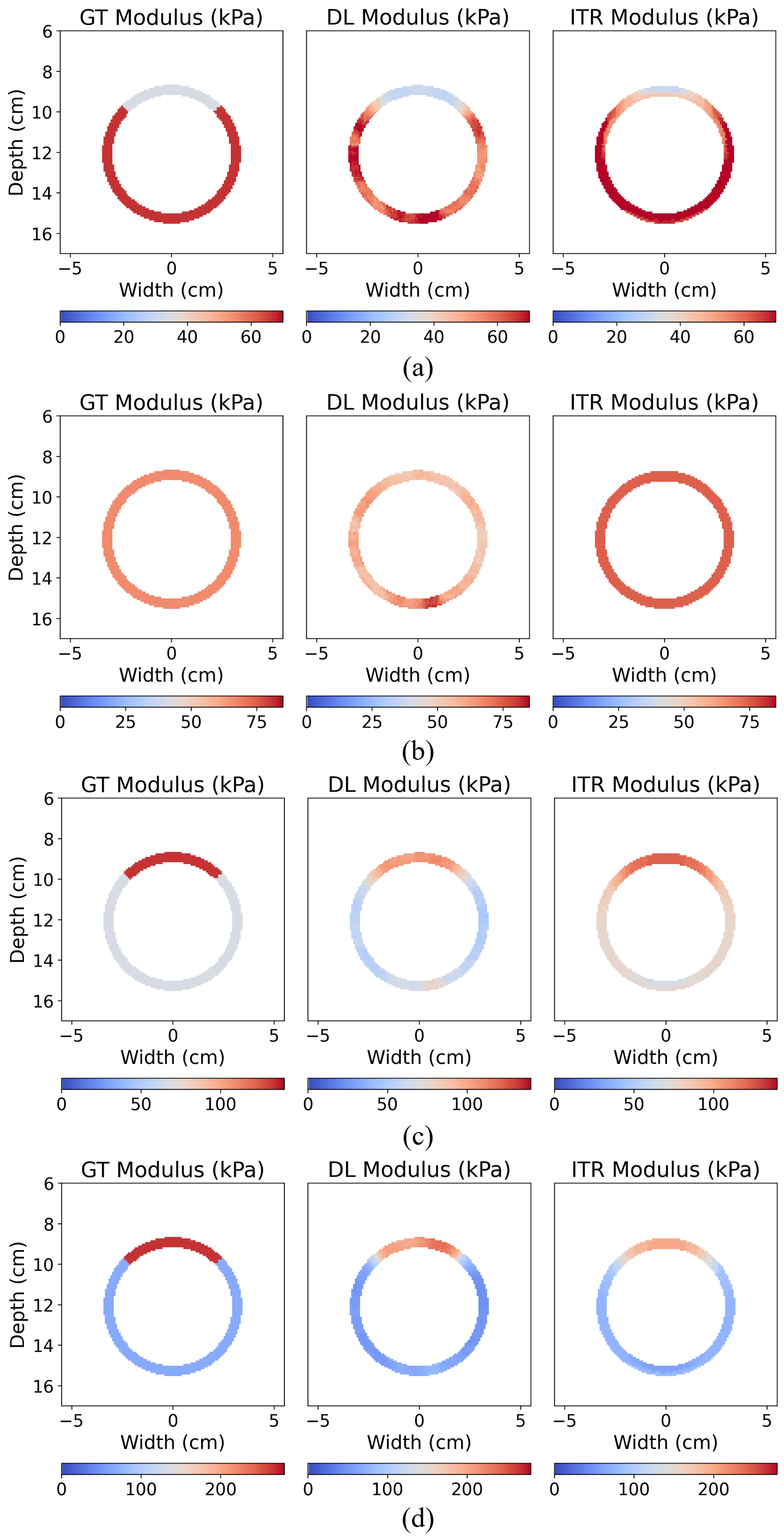}
\caption{COMSOL simulation results for (a) COMSOL model 1, (b) COMSOL model 2, (c) COMSOL model 3, and (d) COMSOL model 4, showing ground truth modulus ($\mu_{ij}$) and predictions from the DL ($\mu_{ij}^{dl}$) and ITR ($\mu_{ij}^{itr}$) reconstruction methods.}
\label{fig:comsol_all}
\end{figure} 

For our digital phantoms simulated from COMSOL, we first calculated the displacement fields from 3D simulated ultrasound images using the image registration algorithm as described in section \ref{sec:ir}. We then input the displacement fields into both ITR and DL reconstruction techniques to predict the modulus distribution. Our digital phantoms were designed with two distinct regions, upper (anterior) region and a lower (posterior) region, composed of different material property. Hence, we divided the predictions into four quadrants (upper, lower, right, and left) with the two corner-to-corner diagonals of the image, assuming their intersection approximated the vessel center. In our digital phantoms, the stiffened region consistently occupied the upper quadrant of the vessel, while the lower quadrant remained within the nominal background modulus region. We calculated the modular ratio between the upper and lower regions, using the equation:

\begin{equation}
\label{eq:modratio}
   \text{Modular Ratio} (\eta) = \frac{ \text{Modulus of upper region}}{\text{Modulus of lower region}}.
\end{equation}

{\setlength{\tabcolsep}{5pt}
\begin{table}[h]
\captionsetup{font=small}
\caption{Summary of evaluation metrics and modulus values on different regions for various COMSOL simulated phantoms.}
\label{tab:results_com_key_metrics}
\centering
\begin{tabular}{c|c|c|c|c}
\hline
\textbf{Metric} & \textbf{Comsol 1} & \textbf{Comsol 2} & \textbf{Comsol 3} & \textbf{Comsol 4}\\
\hline
DL NMSE      & $0.04 $   & $0.06$   & $0.08$   & $0.12$ \\
ITR NMSE        & $0.09 $   & $0.09$   & $0.14$   & $0.22$ \\
DL DSC        & $0.98$   & $0.98$   & $0.98$   & $0.98$ \\
ITR DSC         & $0.96$   & $0.96$   & $0.96$   & $0.96$ \\
\hline
\boldsymbol{$\mu_{\text{up}}$} (kPa)        & $33.3 $   & $66.7$   & $133.3$   & $266.7$ \\
DL-\boldsymbol{$\hat{\mu}_{\text{up}}$} (kPa)   &$30.3\!\pm\!2.2$& $54.2\!\pm\!1.9$&$104.7\!\pm\! 3.9$&$212.0\!\pm\!14.9$\\
ITR-\boldsymbol{$\hat{\mu}_{\text{up}}$} (kPa)    & $39.8 \!\pm\! 6.1$  & $74.2 \!\pm\! 0.2$  & $118.2 \!\pm\! 1.1$  & $186.5 \!\pm\! 14.7$ \\
\boldsymbol{$\mu_{\text{low}}$} (kPa)         & $66.7$  & $66.7$  & $66.7$  & $66.7$  \\
DL-\boldsymbol{$\hat{\mu}_{\text{low}}$} (kPa)   & $63.4 \!\pm\! 7.8$ & $63.1 \!\pm\! 7.4$ & $63.6 \!\pm\! 7.3$ & $63.9 \!\pm\! 7.8$\\
ITR-\boldsymbol{$\hat{\mu}_{\text{low}}$} (kPa)   & $79.3 \!\pm\! 9$ & $74.2 \!\pm\! 0.2$ & $72.9 \!\pm\! 3.8$ & $71.8 \!\pm\! 8.3$ \\
\boldsymbol{$\eta$}           & $0.50$ & $1.00$ & $2.00$ & $4.00$  \\
DL-\boldsymbol{$\hat{\eta}$}     & $0.48$ & $0.86$ & $1.64$ & $3.32$  \\
ITR-\boldsymbol{$\hat{\eta}$}     & $0.50$ & $1.00$ & $1.62$ & $2.59$  \\
\hline
\end{tabular}
\end{table}
}

Table \ref{tab:results_com_key_metrics} provides the details of the evaluation metrics we used for the COMSOL data for both the DL and ITR methods. It reports the NMSE and DSC scores for both methods. Additionally, the modulus values for the upper and lower regions of the four different digital phantoms are summarized in this table where we report the mean and standard deviation for the expected modulus of the upper region (\boldsymbol{$\mu_{\text{up}}$}) and it's predicted modulus with DL (DL-\boldsymbol{$\hat{\mu}_{\text{up}}$}) and ITR (ITR-\boldsymbol{$\hat{\mu}_{\text{up}}$}) method, expected modulus of the lower region (\boldsymbol{$\mu_{\text{low}}$}) and it's predicted modulus with DL (DL-\boldsymbol{$\hat{\mu}_{\text{low}}$}) and ITR (ITR-\boldsymbol{$\hat{\mu}_{\text{low}}$}) method, expected (\boldsymbol{$\eta$}) and predicted modular ratios from DL (DL-\boldsymbol{$\hat{\mu}_{\text{low}}$}) and ITR (ITR-\boldsymbol{$\hat{\mu}_{\text{low}}$}) method. The images of the displacement fields and the predicted modulus distribution are shown in Figure \ref{fig:comsol_all}.

\subsection{Phantom Data}

\begin{figure}[h]
    \centering
    \includegraphics[width=0.99\linewidth]{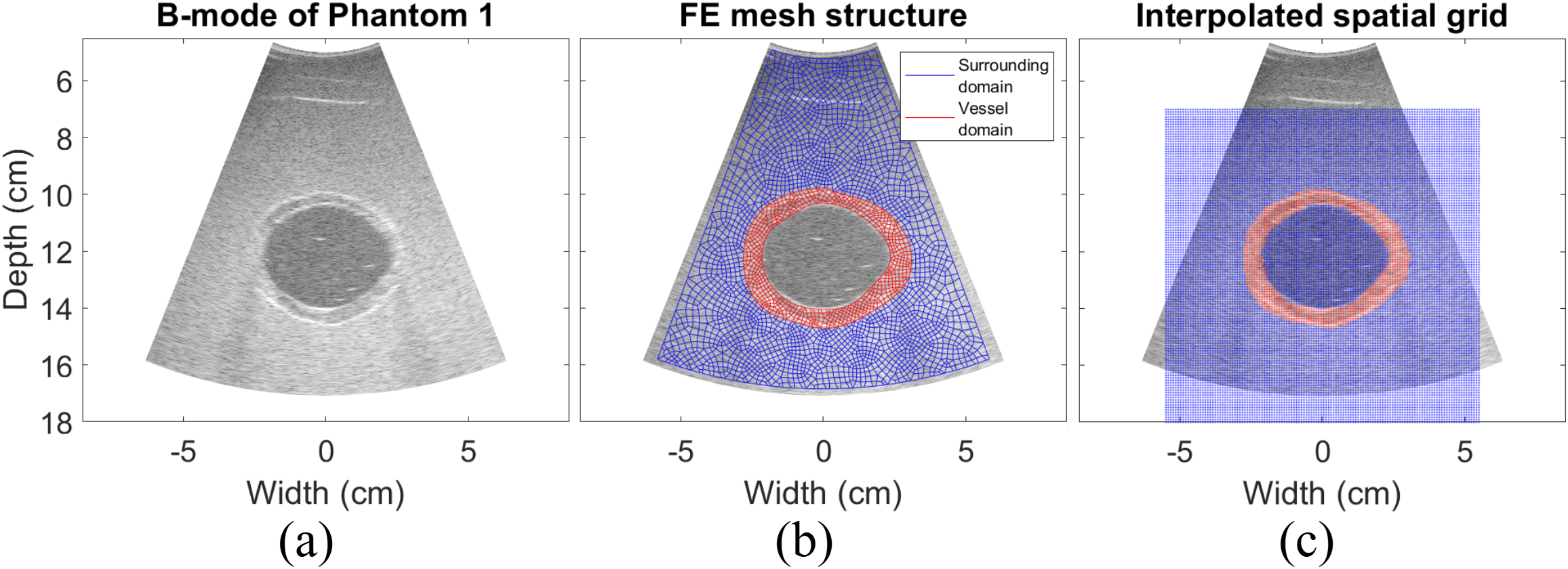}
    \caption{(a) B-mode image of Phantom 1, (b) FE mesh structure used for modeling, and (c) the interpolated spatial grid used as input to the DL method.}
    \label{fig:phantom_mesh}
\end{figure}

For our experimental phantoms, we first calculate the displacement fields from ultrasound RF images using the image registration algorithm, as described in Section \ref{sec:ir}. We then input the pressure-normalized displacement fields into both our ITR algorithm and the DL model and predict the modulus distribution. We divided the predicted image into four quadrants (upper, lower, right, and left) using the two corner-to-corner diagonals of the image. We assumed that the intersecting point of these diagonals coincided approximately with the center of the vessel. The stiffened region of our phantoms always covered the segmented upper quadrant of the vessel image, while the segmented lower quadrant vessel was within the nominal background modulus region. We then calculated the modular ratio between the two regions using equation (\ref{eq:modratio}).

{\setlength{\tabcolsep}{4pt}
\begin{table}[ht]
\captionsetup{font=small}
\caption{Summary of modulus values on different regions for various experimental phantoms.} 
\label{tab:results_pha}
\centering
\begin{tabular}{c|c|c|c|c}
\hline
\textbf{Metric} & \textbf{Phantom 1} & \textbf{Phantom 2} & \textbf{Phantom 3} & \textbf{Phantom 4}\\
\hline
\boldsymbol{$\mu_{\text{up}}$} (kPa)          & $17.4 \!\pm\! 1$   & $48.3 \!\pm\! 5.7$   & $95.1 \!\pm\! 0.4$   & $170 \!\pm\! 4.1$ \\
DL-\boldsymbol{$\hat{\mu}_{\text{up}}$} (kPa)    & $34.1 \!\pm\! 10.8$  & $85.3 \!\pm\! 18.1$  & $126.7 \!\pm\! 13.5$  & $212.3\!\pm\! 10.5$ \\
ITR-\boldsymbol{$\hat{\mu}_{\text{up}}$} (kPa)    & $37.6 \!\pm\! 6$  & $80.5 \!\pm\! 16.3$  & $129.4 \!\pm\! 40.1$  & $154.2 \!\pm\! 83.4$ \\

\boldsymbol{$\mu_{\text{low}}$} (kPa)         & $17.4 \!\pm\! 1$ & $17.4 \!\pm\! 1$ & $17.4 \!\pm\! 1$ & $17.4 \!\pm\! 1$ \\
DL-\boldsymbol{$\hat{\mu}_{\text{low}}$} (kPa)   & $31.1 \!\pm\! 3.5$ & $34.4 \!\pm\! 2.5$ & $25.1 \!\pm\! 2.8$ & $19 \!\pm\! 1.8$\\
ITR-\boldsymbol{$\hat{\mu}_{\text{low}}$} (kPa)   & $33.1 \!\pm\! 4.4$ & $34.1 \!\pm\! 6.9$ & $30.4 \!\pm\! 10.7$ & $21.3 \!\pm\! 7.5$\\
\boldsymbol{$\eta$}           & $1.00$ & $2.78$ & $5.47$ & $9.77$  \\
DL-\boldsymbol{$\hat{\eta}$}     & $1.09$ & $2.48$ & $5.05$ & $11.17$  \\
ITR-\boldsymbol{$\hat{\eta}$}     & $1.13$ & $2.36$ & $4.26$ & $7.26$  \\
\hline
\end{tabular}
\end{table}
}

\begin{figure}[ht!]
\centering
\includegraphics[width=1.0\linewidth]{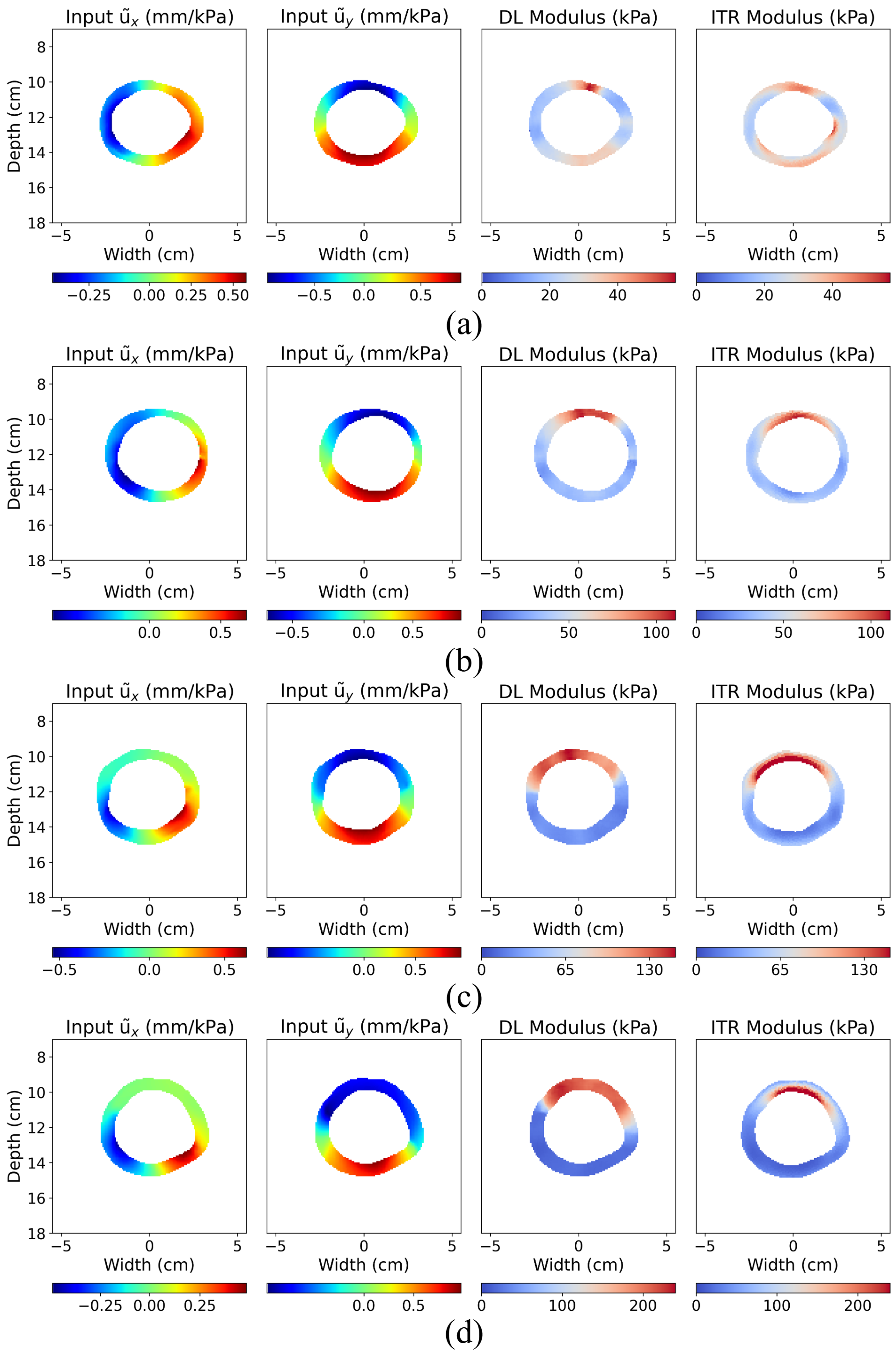}
\caption{Results for (a) phantom 1, (b) phantom 2, (c) phantom 3, and (d) phantom 4, showing pressure-normalized X and Y measured displacement inputs and the corresponding modulus predictions from the DL ($\mu_{ij}^{dl}$) and ITR ($\mu_{ij}^{itr}$) reconstruction methods.}
\label{fig:phantom_all}
\end{figure}

The results for the four different phantoms are summarized in Table \ref{tab:results_pha}. 
In this table, we report the mean and standard deviation for the expected modulus of the upper region (\boldmath{$\mu_{\text{up}}$}) and it's predicted modulus with DL (\boldmath{DL-$\hat{\mu}_{\text{up}}$}) and ITR (\boldmath{ITR-$\hat{\mu}_{\text{up}}$}) method, expected modulus of the lower region (\boldmath{$\mu_{\text{low}}$}) and it's predicted modulus with DL (\boldmath{DL-$\hat{\mu}_{\text{low}}$}) and ITR (\boldmath{ITR-$\hat{\mu}_{\text{low}}$}) method, expected (\boldmath{$\eta$}) and predicted modular ratios from DL (\boldmath{DL-$\hat{\mu}_{\text{low}}$}) and ITR (\boldmath{ITR-$\hat{\mu}_{\text{low}}$}) method. The images of the displacement fields and the predicted modulus distribution are shown in Figure \ref{fig:phantom_all}.

\subsection{Clinical Data}

\begin{figure}[h]
    \centering
    \includegraphics[width=1.0\linewidth]{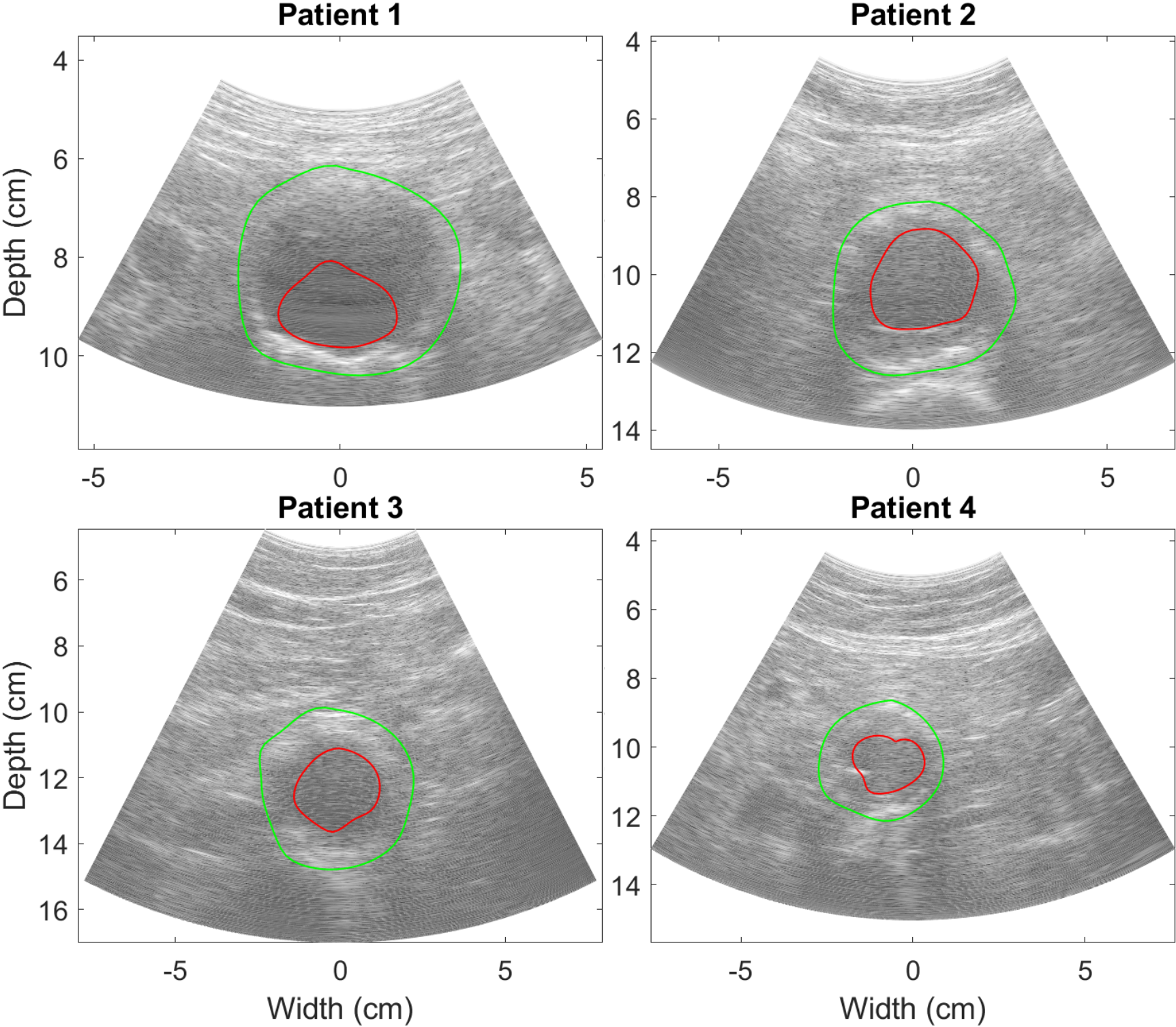}
    \caption{B-mode images of clinical patients with annotation of the aorta location.}
    \label{fig:clinicalbmodes}
\end{figure}

\begin{figure}[t!]
    \centering
    \includegraphics[width=1.0\linewidth]{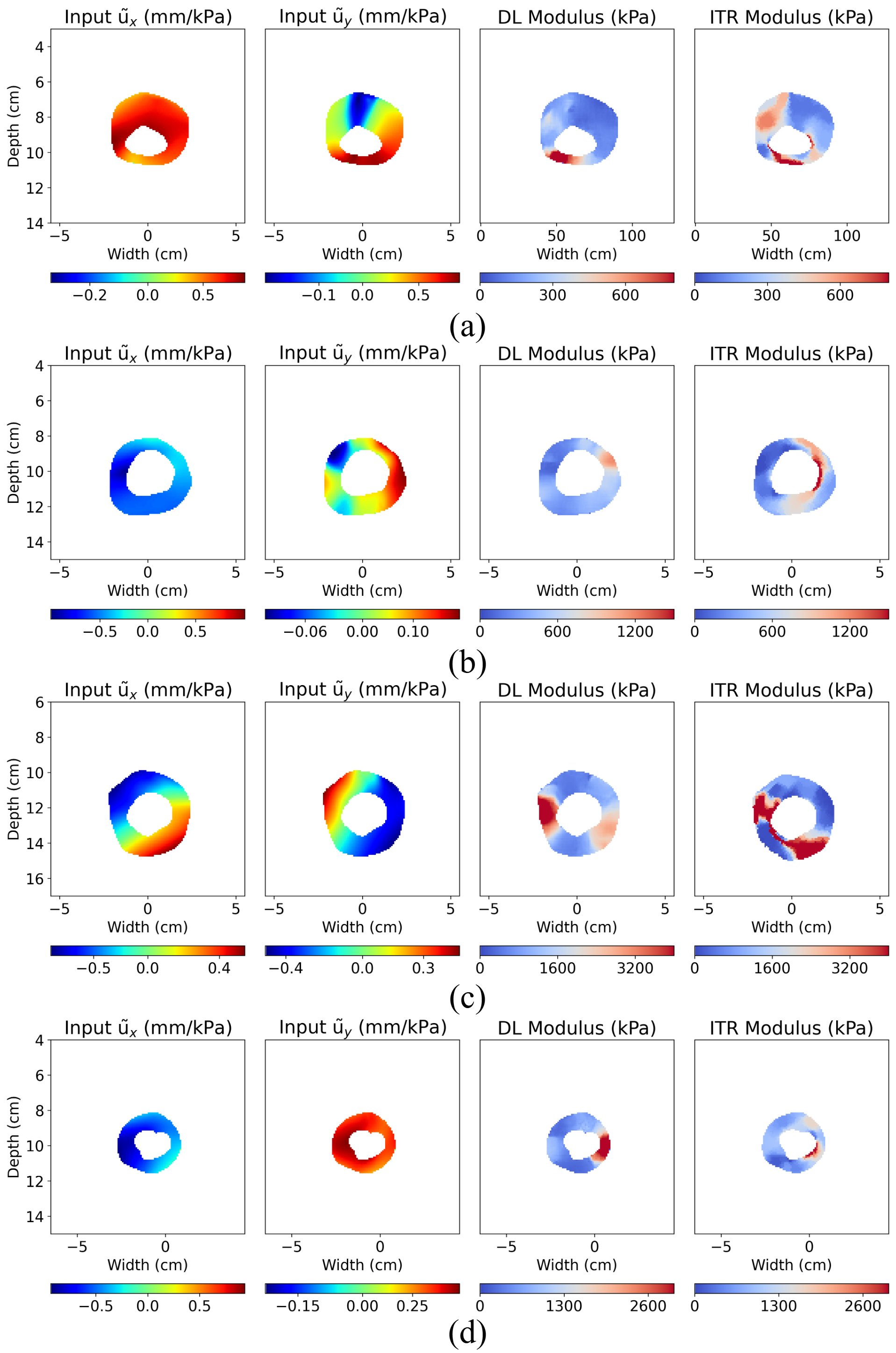}
    \caption{Results for clinical patient (a) 1, (b) 2, (c) 3, and (d) 4, showing pressure-normalized X and Y measured displacement inputs and the corresponding modulus predictions from the DL ($\mu_{ij}^{dl}$) and ITR ($\mu_{ij}^{itr}$) reconstruction methods.}
    \label{fig:results_patient1}
\end{figure}

We applied our iterative algorithm and DL model to four patients collected from our previous clinical study. The displacement fields were obtained from US image sequences using the image registration algorithm, as described above. 
The B-mode images are shown in Figure \ref{fig:clinicalbmodes}, which also depict segmented vessels, blood lumen in red, and outer vessel in green, found by an expert surgeon \cite{zottola2023intermediate}. Upon applying our deep learning and iterative reconstruction method, we obtain the modulus reconstruction as shown in Figure \ref{fig:results_patient1}.


\section{Discussion}
In this study, we created a comprehensive simulated data set of the 2D aorta that we used to train a DL model, which was then applied to both phantom and clinical data. We used pressure-normalized displacements as input to the DL model, inspired by the use of pressure-normalized strain proposed by Mix et al. \cite{mix2017detecting}, to allow quantitative calculation of the modulus. The modulus distributions predicted by our DL model from the axial and lateral displacement fields demonstrate strong performance across key evaluation metrics. The low NMSE in the predicted modulus values in the test set validates the quantitative accuracy of the model. Similarly, the high DSC indicates strong spatial agreement between the predicted and ground truth vessel regions. These results confirm that the chosen architecture is well suited for solving the inverse problem in elastography and effectively captures the underlying mechanical patterns in the data.




For the COMSOL simulated digital phantoms, we obtained modulus values from both the DL model and the ITR algorithm. Overall, both approaches show strong agreement with the ground truth across all four cases. The NMSE values indicate that the DL method consistently achieves lower errors than the ITR method, suggesting that the DL approach generalizes more effectively from 2D training data to the 3D test data and provides more accurate average modulus estimates. Similarly, high DSC score for both methods demonstrate that they can reliably reconstruct the geometry of the vessel with strong overlap relative to the ground truth. 

When comparing the modulus values from the upper and lower regions for the COMSOL data, both methods provide values close to the expected regions. The DL method tended to accurately predict the magnitude of the lower modulus but underestimated the recovered contrast of the upper region relative to the lower. This underestimation may arise because, even in a forward simulation, when localized modulus contrast is present in a modulus distribution, the differences in the displacement fields as the contrast is increased becomes increasingly small and easily obscured by noise. We hypothesize that the DL model may err on the side of reduced contrast when faced with real data. In addition, the momentum-based regularization used here effectively penalizes large jumps in functions of the strains, which may also have reduced the apparent contrast in the modulus prior to being input into the DL and ITR methods.  However, the ITR method overall overestimates the value of the modulus and underestimates the contrast in the modulus to a greater extent than the DL method. This is likely due to the added effect of regularization of the modulus in the optimization cost function of the ITR method. The result is a slight overestimation of the upper quadrant in the COMSOL 1 model and an underestimation of the modulus for the COMSOL 3 and 4 models. With respect to the modular ratio, the DL method achieves better estimates at higher contrasts, whereas the ITR method performs better at lower contrasts, possibly for similar reasons. It is also likely that the loss of contrast recovery in the ITR technique could have affected the overall magnitude of the modulus. The algorithm may have adjusted the average modulus, rather than the contrast penalized in the regularization, to best match the displacement measurements. However, both approaches capture the contrast ratio with reasonable accuracy, which is itself a valuable biomarker \cite{dongvivo}.

Interestingly, for the phantom data, the DL method and the ITR method show reasonable agreement for most experiments with phantoms (1-3). Both methods tend to overestimate modulus values in all cases. Specifically for Phantom 4, ITR method tends to underestimate the expected modulus contrast, based on the independent measurements, probably for the same reasons as observed in the COMSOL generated data. Moreover, DL provides more accurate contrast ratio estimates compared to ITR, which often exaggerates contrast at higher values. The reasons for these offsets require further study, but both methods nonetheless preserve the relative contrast between regions.

For ITR reconstructions, a noticeable radial variation was observed in the COMSOL predictions and, more prominently, in the phantom predictions. This artifact could be likely be attributed to the nature of the optimization and the regularization strategy. Clearly, iterative optimization had a bias toward increasing the magnitude of the modulus closer to the vessel lumen. However, such patterns do not appear in the DL reconstruction, likely due to the fact that radial variations were not present in the simulated data used for training.

While the ITR approach is based on explicit regularization terms to ensure stability and converge to an optimized solution, the DL model learns from the large amount of data. Although the foundational approaches differ, one rooted in a physics-based model, while the other is data-driven, both methods yielded similar results. Incorporating physics into the deep learning framework could further enhance its performance, which could be a direction for future research.

\begin{figure}[h]
    \centering
    \includegraphics[width=1.0\linewidth]{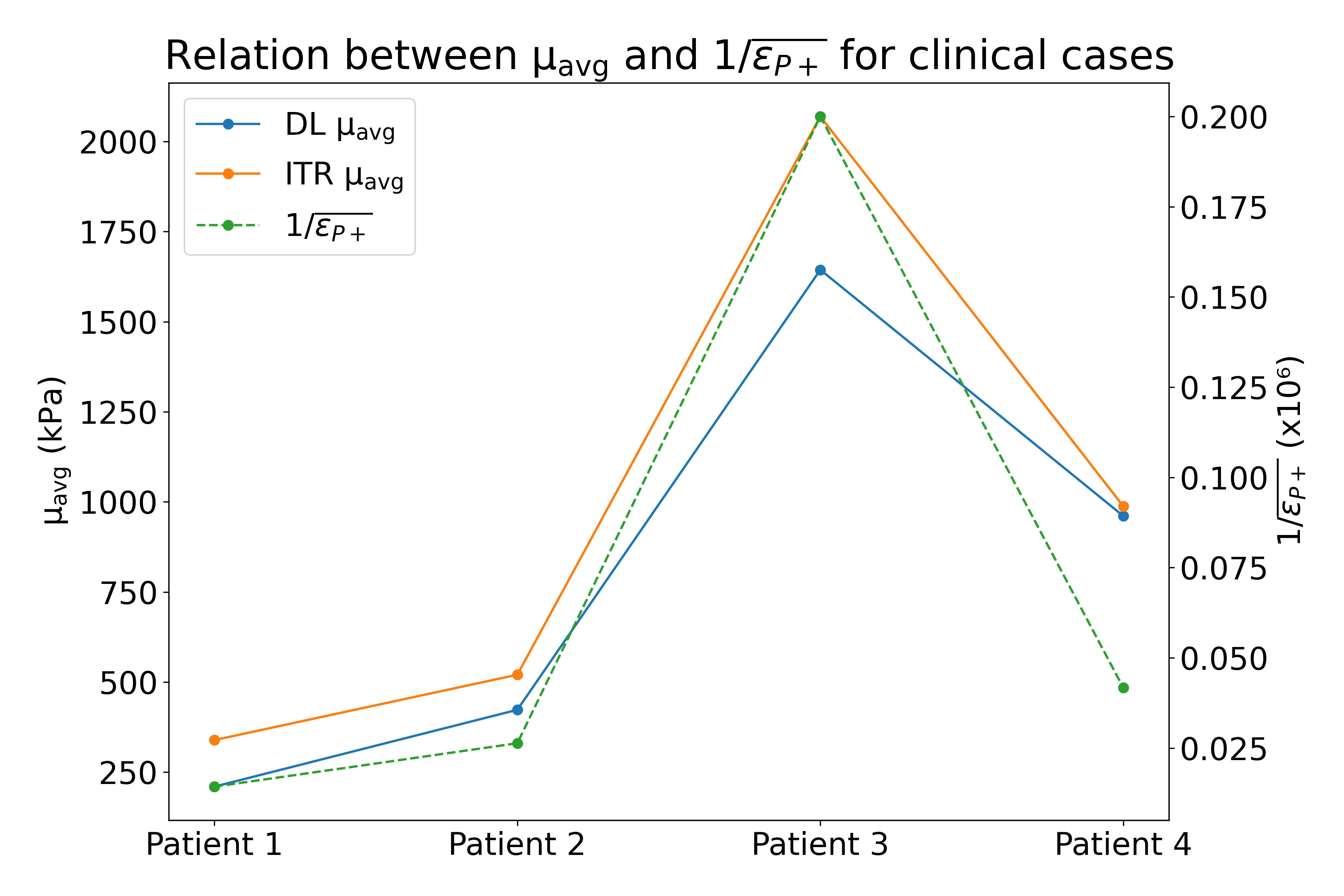}
    \caption{Plot showing the relationship between the average modulus and the inverse of the average pressure normalized maximum principal strain across different clinical patients.}
    \label{fig:strainanalysis}
\end{figure}

For the clinical data, where ground truth modulus values were not available, we performed a strain-based analysis of the measured displacement data. For each clinical case, we calculate the maximum principal strain distribution from the estimated displacement fields. From this we calculate the average maximum principal strain, within the vessel, and normalize to arrive at the pressure normalized maximum principal strain ($\overline{\varepsilon_{P+}}$), similar to that calculated in Mix et. al. \cite{mix2017detecting}. 
We also calculated the average of predicted modulus ($\mu_{\text{avg}}$) for each clinical case. 
Theoretically, $\overline{\varepsilon_{P+}}$, which is an estimate of vessel compliance, and $\mu_{\text{avg}}$ are expected to exhibit an inverse relationship between the images. This expected trend is indeed observed in Figure \ref{fig:strainanalysis}, where both the DL and ITR estimated values show similar average modulus values and are seemingly correlated to $1/\overline{\varepsilon_{P+}}$, supporting the validity of our predictions even in the absence of ground truth values.



The computational efficiency of the DL method demonstrates a clear advantage over the ITR approach. The processing time for the DL prediction was on average 0.058 seconds per example, compared to 158.57 seconds for the ITR method. The image registration took an average of 55.02 seconds per example, likely due to the accumulation process, which could be reduced with more optimized algorithm. This highlights the effectiveness of DL as a powerful tool for modulus reconstruction, with strong potential for real-time application in the future.

One of the limitations of this study is that the simulated training data was restricted to 2D. Although it would be possible to generate 3D data, as in the COMSOL simulations, this approach is highly time consuming and computationally expensive. In the future, we plan to investigate simplified 3D models that may simulate a 3D deforming material, but with limited geometries or using simplified BCs such that generating large data sets is computationally tractable. Nevertheless, even with 2D data, the reconstructions obtained here demonstrate strong agreement with the ground truth. 
Another limitation lies in the nature of the simulated displacement fields used for training, which are noiseless except for minor discretization artifacts. However, displacement fields measured from ultrasound data are affected by multiple sources of error, including discretization inaccuracies, limitations in displacement estimation algorithms, and inherent noise in the raw RF signals. These differences between simulated and real-world data may impact model generalization, particularly in experimental and clinical applications. In future work, we intend to conduct noise and sensitivity studies for these methods. 
The present work primarily demonstrates that a deep learning framework can be applied to this problem and achieve performance comparable to the ITR method. In the future, our goal is to improve the DL model through exploration of various architectures, hyperparameter tuning, and the generation of more realistic simulated ultrasound data.
Currently, we require medical experts to select the ROI and appropriate frames from ultrasound sequences manually. We plan to automate this process and incorporate a greater number of US frames during training to reduce the dependency on manual selection. We also expect that the local overall accuracy of our modulus reconstructions will increase by using frames over multiple cardiac cycles to estimate the modulus.
Finally, we plan to expand our clinical data set and perform more extensive \textit{in vivo} evaluations to rigorously test and validate the practical applicability of the method.

\section{Conclusion}
We demonstrate that a deep learning model, trained only on FE simulated data, can be used to predict shear modulus from both simulated and experimental ultrasound elastography displacement estimates. Our study includes model training using simulated displacement fields, followed by validation on digital phantoms, experimental phantoms, and preliminary clinical data. The results show that the model can generalize well across different variations and successfully learn the relationship between displacement fields and modulus distribution. Moreover, we compared the deep learning model with iterative reconstruction methods which shows comparable performance. This suggests that, with sufficiently diverse and representative simulated training data of AAAs, deep learning models can generalize well to real-world experimental conditions for estimating the shear modulus of the aorta. The clinical results provide encouraging evidence for the applicability of our method in practical settings.

\ifCLASSOPTIONcaptionsoff
  \newpage
\fi



%



\bibliographystyle{IEEEtran}
\bibliography{main}

\begin{thebibliography}{10}
\providecommand{\url}[1]{#1}
\csname url@samestyle\endcsname
\providecommand{\newblock}{\relax}
\providecommand{\bibinfo}[2]{#2}
\providecommand{\BIBentrySTDinterwordspacing}{\spaceskip=0pt\relax}
\providecommand{\BIBentryALTinterwordstretchfactor}{4}
\providecommand{\BIBentryALTinterwordspacing}{\spaceskip=\fontdimen2\font plus
\BIBentryALTinterwordstretchfactor\fontdimen3\font minus \fontdimen4\font\relax}
\providecommand{\BIBforeignlanguage}[2]{{%
\expandafter\ifx\csname l@#1\endcsname\relax
\typeout{** WARNING: IEEEtran.bst: No hyphenation pattern has been}%
\typeout{** loaded for the language `#1'. Using the pattern for}%
\typeout{** the default language instead.}%
\else
\language=\csname l@#1\endcsname
\fi
#2}}
\providecommand{\BIBdecl}{\relax}
\BIBdecl

\bibitem{aggarwal2011abdominal}
S.~Aggarwal, A.~Qamar, V.~Sharma, and A.~Sharma, ``Abdominal aortic aneurysm: A comprehensive review,'' \emph{Experimental \& Clinical Cardiology}, vol.~16, no.~1, p.~11, 2011.

\bibitem{pande2008abdominal}
R.~L. Pande and J.~A. Beckman, ``Abdominal aortic aneurysm: populations at risk and how to screen,'' \emph{Journal of Vascular and Interventional Radiology}, vol.~19, no.~6, pp. S2--S8, 2008.

\bibitem{kent2010analysis}
K.~C. Kent, R.~M. Zwolak, N.~N. Egorova, T.~S. Riles, A.~Manganaro, A.~J. Moskowitz, A.~C. Gelijns, and G.~Greco, ``Analysis of risk factors for abdominal aortic aneurysm in a cohort of more than 3 million individuals,'' \emph{Journal of vascular surgery}, vol.~52, no.~3, pp. 539--548, 2010.

\bibitem{fleming2005screening}
C.~Fleming, E.~P. Whitlock, T.~L. Beil, and F.~A. Lederle, ``Screening for abdominal aortic aneurysm: a best-evidence systematic review for the us preventive services task force,'' \emph{Annals of internal medicine}, vol. 142, no.~3, pp. 203--211, 2005.

\bibitem{benson2018ultrasound}
R.~A. Benson, L.~Meecham, O.~Fisher, and I.~M. Loftus, ``Ultrasound screening for abdominal aortic aneurysm: current practice, challenges and controversies,'' \emph{The British journal of radiology}, vol.~91, no. 1090, p. 20170306, 2018.

\bibitem{lindholt1999validity}
J.~Lindholt, S.~Vammen, S.~Juul, E.~Henneberg, and H.~Fasting, ``The validity of ultrasonographic scanning as screening method for abdominal aortic aneurysm,'' \emph{European journal of vascular and endovascular surgery}, vol.~17, no.~6, pp. 472--475, 1999.

\bibitem{sprouse2004ultrasound}
L.~Sprouse~Ii, G.~Meier~III, F.~Parent, R.~DeMasi, M.~Glickman, and G.~Barber, ``Is ultrasound more accurate than axial computed tomography for determination of maximal abdominal aortic aneurysm diameter?'' \emph{European journal of vascular and endovascular surgery}, vol.~28, no.~1, pp. 28--35, 2004.

\bibitem{brewster2003guidelines}
D.~C. Brewster, J.~L. Cronenwett, J.~W. Hallett~Jr, K.~W. Johnston, W.~C. Krupski, and J.~S. Matsumura, ``Guidelines for the treatment of abdominal aortic aneurysms: report of a subcommittee of the joint council of the american association for vascular surgery and society for vascular surgery,'' \emph{Journal of vascular surgery}, vol.~37, no.~5, pp. 1106--1117, 2003.

\bibitem{khan2015assessing}
S.~Khan, V.~Verma, S.~Verma, S.~Polzer, and S.~Jha, ``Assessing the potential risk of rupture of abdominal aortic aneurysms,'' \emph{Clinical radiology}, vol.~70, no.~1, pp. 11--20, 2015.

\bibitem{nordon2009review}
I.~M. Nordon, R.~J. Hinchliffe, P.~J. Holt, I.~M. Loftus, and M.~M. Thompson, ``Review of current theories for abdominal aortic aneurysm pathogenesis,'' \emph{Vascular}, vol.~17, no.~5, pp. 253--263, 2009.

\bibitem{VORP20071887}
\BIBentryALTinterwordspacing
D.~A. Vorp, ``Biomechanics of abdominal aortic aneurysm,'' \emph{Journal of Biomechanics}, vol.~40, no.~9, pp. 1887--1902, 2007. [Online]. Available: \url{https://www.sciencedirect.com/science/article/pii/S002192900600323X}
\BIBentrySTDinterwordspacing

\bibitem{vorp2005biomechanical}
D.~A. Vorp and J.~P.~V. Geest, ``Biomechanical determinants of abdominal aortic aneurysm rupture,'' \emph{Arteriosclerosis, thrombosis, and vascular biology}, vol.~25, no.~8, pp. 1558--1566, 2005.

\bibitem{gasser2016biomechanical}
T.~C. Gasser, ``Biomechanical rupture risk assessment,'' \emph{Aorta}, vol.~4, no.~02, pp. 42--60, 2016.

\bibitem{kadoglou2012arterial}
N.~P. Kadoglou, I.~Papadakis, K.~G. Moulakakis, I.~Ikonomidis, M.~Alepaki, P.~Moustardas, S.~Lampropoulos, P.~Karakitsos, J.~Lekakis, and C.~D. Liapis, ``Arterial stiffness and novel biomarkers in patients with abdominal aortic aneurysms,'' \emph{Regulatory peptides}, vol. 179, no. 1-3, pp. 50--54, 2012.

\bibitem{sonesson1999abdominal}
B.~Sonesson, T.~Sandgren, and T.~L{\"a}nne, ``Abdominal aortic aneurysm wall mechanics and their relation to risk of rupture,'' \emph{European Journal of Vascular and Endovascular Surgery}, vol.~18, no.~6, pp. 487--493, 1999.

\bibitem{gennisson2013ultrasound}
J.-L. Gennisson, T.~Deffieux, M.~Fink, and M.~Tanter, ``Ultrasound elastography: principles and techniques,'' \emph{Diagnostic and interventional imaging}, vol.~94, no.~5, pp. 487--495, 2013.

\bibitem{sigrist2017ultrasound}
R.~M. Sigrist, J.~Liau, A.~El~Kaffas, M.~C. Chammas, and J.~K. Willmann, ``Ultrasound elastography: review of techniques and clinical applications,'' \emph{Theranostics}, vol.~7, no.~5, p. 1303, 2017.

\bibitem{richards2011investigating}
M.~Richards and M.~M. Doyley, ``Investigating the impact of spatial priors on the performance of model-based ivus elastography,'' \emph{Physics in Medicine \& Biology}, vol.~56, no.~22, p. 7223, 2011.

\bibitem{tuladhar2023deep}
U.~R. Tuladhar, R.~A. Simon, C.~A. Linte, and M.~S. Richards, ``A deep learning framework to estimate elastic modulus from ultrasound measured displacement fields,'' in \emph{Proceedings of SPIE--the International Society for Optical Engineering}, vol. 12470, 2023, p. 124700P.

\bibitem{fromageau2005feasibility}
J.~Fromageau, S.~Lerouge, G.~Soulez, I.~Salazkin, J.~Raymond, and G.~Cloutier, ``A feasibility study of elastography on abdominal aneurysms,'' in \emph{IEEE Ultrasonics Symposium, 2005.}, vol.~1.\hskip 1em plus 0.5em minus 0.4em\relax IEEE, 2005, pp. 257--260.

\bibitem{bonnardeaux_vivo_2024}
\BIBentryALTinterwordspacing
V.~Bonnardeaux, R.~Moore, M.-H. Roy-Cardinal, V.~Guerrera, E.~Therasse, C.~C. Lefebvre, S.~Amouri, E.~D. Martino, A.~Forneris, G.~Cloutier, and G.~Soulez, ``In vivo assessment of abdominal aortic aneurysm wall strain using 2d and 3d b-mode ultrasound and dynamic {CT} scan,'' vol.~79, no.~6, pp. e250--e251, publisher: Elsevier. [Online]. Available: \url{https://www.jvascsurg.org/article/S0741-5214(24)00814-0/fulltext}
\BIBentrySTDinterwordspacing

\bibitem{trachet2015performance}
B.~Trachet, R.~A. Fraga-Silva, F.~J. Londono, A.~Swillens, N.~Stergiopulos, and P.~Segers, ``Performance comparison of ultrasound-based methods to assess aortic diameter and stiffness in normal and aneurysmal mice,'' \emph{PloS one}, vol.~10, no.~5, p. e0129007, 2015.

\bibitem{voizard_feasibility_2020}
\BIBentryALTinterwordspacing
N.~Voizard, A.~Bertrand-Grenier, H.~Alturkistani, E.~Therasse, A.~Tang, C.~Kauffmann, G.~Cloutier, and G.~Soulez, ``Feasibility of shear wave sonoelastography to detect endoleak and evaluate thrombus organization after endovascular repair of abdominal aortic aneurysm,'' vol.~30, no.~7, pp. 3879--3889. [Online]. Available: \url{https://doi.org/10.1007/s00330-020-06739-3}
\BIBentrySTDinterwordspacing

\bibitem{mix2017detecting}
D.~S. Mix, L.~Yang, C.~C. Johnson, N.~Couper, B.~Zarras, I.~Arabadjis, L.~E. Trakimas, M.~C. Stoner, S.~W. Day, and M.~S. Richards, ``Detecting regional stiffness changes in aortic aneurysmal geometries using pressure-normalized strain,'' \emph{Ultrasound in medicine \& biology}, vol.~43, no.~10, pp. 2372--2394, 2017.

\bibitem{van2019quantification}
E.~M. van Disseldorp, N.~J. Petterson, F.~N. van~de Vosse, M.~R. van Sambeek, and R.~G. Lopata, ``Quantification of aortic stiffness and wall stress in healthy volunteers and abdominal aortic aneurysm patients using time-resolved 3d ultrasound: a comparison study,'' \emph{European Heart Journal-Cardiovascular Imaging}, vol.~20, no.~2, pp. 185--191, 2019.

\bibitem{long2005compliance}
A.~Long, L.~Rouet, A.~Bissery, P.~Rossignol, D.~Mouradian, and M.~Sapoval, ``Compliance of abdominal aortic aneurysms evaluated by tissue doppler imaging: correlation with aneurysm size,'' \emph{Journal of vascular surgery}, vol.~42, no.~1, pp. 18--26, 2005.

\bibitem{isola2017image}
P.~Isola, J.-Y. Zhu, T.~Zhou, and A.~A. Efros, ``Image-to-image translation with conditional adversarial networks,'' in \emph{Proceedings of the IEEE conference on computer vision and pattern recognition}, 2017, pp. 1125--1134.

\bibitem{li2022deep}
H.~Li, M.~Bhatt, Z.~Qu, S.~Zhang, M.~C. Hartel, A.~Khademhosseini, and G.~Cloutier, ``Deep learning in ultrasound elastography imaging: A review,'' \emph{Medical Physics}, 2022.

\bibitem{wu2018direct}
S.~Wu, Z.~Gao, Z.~Liu, J.~Luo, H.~Zhang, and S.~Li, ``Direct reconstruction of ultrasound elastography using an end-to-end deep neural network,'' in \emph{International conference on medical image computing and computer-assisted intervention}.\hskip 1em plus 0.5em minus 0.4em\relax Springer, 2018, pp. 374--382.

\bibitem{gao2019learning}
Z.~Gao, S.~Wu, Z.~Liu, J.~Luo, H.~Zhang, M.~Gong, and S.~Li, ``Learning the implicit strain reconstruction in ultrasound elastography using privileged information,'' \emph{Medical image analysis}, vol.~58, p. 101534, 2019.

\bibitem{tuladhar2025ultrasound}
U.~R. Tuladhar, R.~A. Simon, C.~A. Linte, and M.~S. Richards, ``Ultrasound elastic modulus reconstruction using a deep learning model trained with simulated data,'' \emph{Journal of Medical Imaging}, vol.~12, no.~1, pp. 017\,001--017\,001, 2025.

\bibitem{lu2019deepaaa}
J.-T. Lu, R.~Brooks, S.~Hahn, J.~Chen, V.~Buch, G.~Kotecha, K.~P. Andriole, B.~Ghoshhajra, J.~Pinto, P.~Vozila \emph{et~al.}, ``Deepaaa: clinically applicable and generalizable detection of abdominal aortic aneurysm using deep learning,'' in \emph{Medical Image Computing and Computer Assisted Intervention--MICCAI 2019: 22nd International Conference, Shenzhen, China, October 13--17, 2019, Proceedings, Part II 22}.\hskip 1em plus 0.5em minus 0.4em\relax Springer, 2019, pp. 723--731.

\bibitem{hong2016automatic}
H.~A. Hong and U.~Sheikh, ``Automatic detection, segmentation and classification of abdominal aortic aneurysm using deep learning,'' in \emph{2016 IEEE 12th International Colloquium on Signal Processing \& Its Applications (CSPA)}.\hskip 1em plus 0.5em minus 0.4em\relax IEEE, 2016, pp. 242--246.

\bibitem{maas2024automatic}
E.~J. Maas, N.~Awasthi, E.~G. van Pelt, M.~R. van Sambeek, and R.~G. Lopata, ``Automatic segmentation of abdominal aortic aneurysms from time-resolved 3-d ultrasound images using deep learning,'' \emph{IEEE Transactions on Ultrasonics, Ferroelectrics, and Frequency Control}, vol.~71, no.~11, pp. 1420--1428, 2024.

\bibitem{vaitenas2023abdominal}
G.~Vait{\.e}nas, V.~Mosenko, A.~Ra{\v{c}}yt{\.e}, K.~Medelis, A.~Skreb{\=u}nas, and T.~Baltr{\=u}nas, ``Abdominal aortic aneurysm diameter versus volume: a systematic review,'' \emph{Biomedicines}, vol.~11, no.~3, p. 941, 2023.

\bibitem{koole2013intraluminal}
D.~Koole, H.~J. Zandvoort, A.~Schoneveld, A.~Vink, J.~A. Vos, L.~L. van~den Hoogen, J.-P.~P. de~Vries, G.~Pasterkamp, F.~L. Moll, and J.~A. van Herwaarden, ``Intraluminal abdominal aortic aneurysm thrombus is associated with disruption of wall integrity,'' \emph{Journal of vascular surgery}, vol.~57, no.~1, pp. 77--83, 2013.

\bibitem{zottola2023intermediate}
Z.~R. Zottola, D.~S. Kong, A.~N. Medhekar, L.~E. Frye, S.~B. Hao, D.~W. Gonring, A.~A. Hirad, M.~C. Stoner, M.~S. Richards, and D.~S. Mix, ``Intermediate pressure-normalized principal wall strain values are associated with increased abdominal aortic aneurysmal growth rates,'' \emph{Frontiers in Cardiovascular Medicine}, vol.~10, p. 1232844, 2023.

\bibitem{GeuzaineRemacle2009}
C.~Geuzaine and J.-F. Remacle, ``{Gmsh: a three-dimensional finite element mesh generator with built-in pre- and post-processing facilities},'' \emph{International Journal for Numerical Methods in Engineering}, vol.~79, no.~11, pp. 1309--1331, 2009.

\bibitem{hughes2003finite}
T.~J. Hughes, \emph{The finite element method: linear static and dynamic finite element analysis}.\hskip 1em plus 0.5em minus 0.4em\relax Courier Corporation, 2003.

\bibitem{sola1997importance}
J.~Sola and J.~Sevilla, ``Importance of input data normalization for the application of neural networks to complex industrial problems,'' \emph{IEEE Transactions on nuclear science}, vol.~44, no.~3, pp. 1464--1468, 1997.

\bibitem{ronneberger2015u}
O.~Ronneberger, P.~Fischer, and T.~Brox, ``U-net: Convolutional networks for biomedical image segmentation,'' in \emph{Medical image computing and computer-assisted intervention--MICCAI 2015: 18th international conference, Munich, Germany, October 5-9, 2015, proceedings, part III 18}.\hskip 1em plus 0.5em minus 0.4em\relax Springer, 2015, pp. 234--241.

\bibitem{dice1945measures}
L.~R. Dice, ``Measures of the amount of ecologic association between species,'' \emph{Ecology}, vol.~26, no.~3, pp. 297--302, 1945.

\bibitem{dongvivo}
H.~Dong, B.~Raterman, M.~Eisner, G.~Brock, R.~D. White, J.~Starr, M.~Haurani, M.~Go, P.~Vaccaro, and A.~Kolipaka, ``In vivo aortic mr elastography in abdominal aortic aneurysm patients: A potential biomarker for predicting aneurysmal events.''

\end{thebibliography}

%








\end{document}